\documentclass[onecolumn,amsmath,showpacs,nofootinbib,12pt]{revtex4}
\usepackage{graphicx}
\usepackage{dcolumn}
\usepackage{bm}
\usepackage{slashed}
\usepackage{float}
\usepackage{mathrsfs}
\usepackage{bm}
\usepackage{float}
\usepackage{fancyhdr}
\usepackage{longtable}
\usepackage{hyperref}
\usepackage{subfigure}
\usepackage{epsfig}
\usepackage{color,soul}
\usepackage{menukeys}
\usepackage{xcolor}
\usepackage{amssymb}
\usepackage{braket}
\usepackage{graphicx}
\usepackage{dcolumn}
\usepackage[mathscr]{eucal}
\usepackage{amsmath}
\usepackage{tabularx}
\usepackage{booktabs}
\usepackage{slashed}
\usepackage{mathtools,slashed}

\allowdisplaybreaks

\makeatletter
\def\pslashed#1{\expandafter\ifx\csname psla@\string#1\endcsname\relax
{\mathpalette{\sla@/00}{\phantom{#1}}}\else
\csname psla@\string#1\endcsname\fi}
\def\declarepslashed#1#2#3#4#5{\expandafter\def\csname psla@\string#5\endcsname{#1{\mathpalette{\sla@{#2}{#3}{#4}}{\phantom{#5}}}}}
\makeatother
\declarepslashed{}{/}{.08}{0}{D}
\newcommand{\be}{\begin{equation}}
\newcommand{\ee}{\end{equation}}
\newcommand{\bea}{\begin{eqnarray}}
\newcommand{\eea}{\end{eqnarray}}
\newcommand{\ben}{\begin{enumerate}}
\newcommand{\een}{\end{enumerate}}
\newcommand{\bde}{\begin{widetext}}
\newcommand{\ede}{\end{widetext}}

\newcommand{\crn}{\nonumber \\}

\newcommand{\al}{\alpha}
\newcommand{\la}{\lambda}

\newcommand{\fr}{\frac}
\newcommand{\bc}{\begin{center}}
\newcommand{\ec}{\end{center}}
\newcommand{\Ga}{\Gamma}
\newcommand{\ga}{\gamma}

\newcommand{\ep}{\epsilon}

\newcommand{\ka}{\kappa}
\newcommand{\La}{\Lambda}

\newcommand{\AdrHEPC}{Phenikaa Institute for Advanced Study, Phenikaa University, Nguyen Trac, Duong Noi, Hanoi 100000, Vietnam}
\newcommand{\AdrH}{Institute of Physics, Vietnam Academy of Science and Technology, 10 Dao Tan, Ba Dinh, Hanoi 100000, Vietnam}
\begin{document}

\title{A Minimal Dark $U(1)_D$ Framework for Inverse Seesaw Neutrino Masses and Dark Matter}
\author{D. T. Huong}
\email{dthuong@iop.vast.vn}
\affiliation{\AdrH} 
\author{N. T. Duy} 
\email{ntduy@iop.vast.vn}
\affiliation{\AdrH} 
\author{Phung Van Dong} 
\email{dong.phungvan@phenikaa-uni.edu.vn}
\affiliation{\AdrHEPC}
\date{\today }
\begin{abstract}
We propose a minimal framework based on a dark $U(1)_D$ gauge symmetry that simultaneously accounts for neutrino masses and dark matter within an inverse seesaw realization. In this setup, the smallness of light neutrino masses is controlled, suppressed by a lepton-number violating parameter $\mu$, which arises dynamically by dark field corrections rather than being introduced by hand. The limit $\mu \to 0$ restores lepton number symmetry, ensuring its, thus neutrino mass, smallness in the sense of ’t Hooft naturalness. We analyze the neutrino mass matrix and active--sterile mixing, highlighting their impact on non-unitarity and charged lepton flavor violation. The model is consistent with current experimental constraints while allowing potentially observable signals, such as $\mu \to e \gamma$. The dark $U(1)_D$ symmetry stabilizes the dark matter candidate and links the neutrino and dark sectors. Viable parameter regions satisfying dark matter relic density, direct detection, and collider bounds are identified. This framework provides a minimal and predictive realization of neutrino mass generation and dark matter stability with a naturally small $\mu$ parameter.
\end{abstract}

\author{}
\maketitle
\section{Introduction}
The observation of neutrino oscillations provides a compelling evidence that neutrinos have nonzero small masses and flavor mixing~\cite{Fukuda:1998mi,Ahmad:2002jz,Esteban:2020cvm}, thereby establishing a clear need for the new physics beyond the Standard Model (SM). Among the various mechanisms proposed so far to account for neutrino masses, the inverse seesaw framework stands out as a particularly appealing scenario, as it naturally accommodates small neutrino masses while keeping the scale of new physics at TeV within experimentally accessible regimes~\cite{Wyler:1982dd,Mohapatra:1986bd}. 
In this framework, the smallness of neutrino masses is governed by a lepton-number violating parameter $\mu$, rather than by the mass scale of the sterile neutrinos. 
As a result, the neutrino mass scale is effectively decoupled from the scale of new physics, allowing for a rich phenomenology.

Despite its phenomenological advantages, the inverse seesaw mechanism raises an important theoretical question regarding the origin and naturalness of the small lepton-number violating parameter $\mu$. 
In many realizations, $\mu$ is introduced by hand as a small Majorana mass term, without a compelling dynamical explanation for its suppressed scale. 
According to the criterion of 't~Hooft naturalness, a parameter is considered as natural smallness only if the symmetry of the theory is enhanced in the limit where it vanishes~\cite{tHooft:1979rat}. 
In the inverse seesaw framework, taking the limit $\mu \to 0$ restores lepton number conservation, thereby technically justifying its smallness. 
Nevertheless, this observation motivates the construction of models in which $\mu$ is not ad hoc inserted, but instead forbidden by a symmetry at tree level and generated dynamically through radiative corrections, for instance~\cite{Ma:2006km,Law:2013gma}. 
Such a dynamical origin for $\mu$ can enhance the predictive possibility of the model and potentially establish a connection between neutrino mass generation and other sectors, such as dark matter (DM).

Abelian gauge extensions of the SM, for instance the baryon-minus-lepton number symmetry $U(1)_{B-L}$, provide a well-motivated framework for neutrino mass generation, as anomaly cancellation requires the presence of right-handed neutrinos~\cite{Marshak:1979fm,Mohapatra:1980qe}. In the minimal realizations, these models typically generate neutrino masses via the type-I seesaw mechanism~\cite{Minkowski:1977sc}, which often involves a high symmetry-breaking scale and thus limits their direct experimental testability. Furthermore, such constructions neither automatically explain the smallness of lepton-number violation by the virtue of 't~Hooft naturalness nor provide a natural connection to dark matter.

On the other hand, extensions of the SM involving an additional $U(1)_D$ gauge symmetry in which the SM particles experience zero $D$-charge have been widely studied as viable frameworks for DM phenomenology. 
In minimal realizations, the corresponding dark gauge boson can interact with the SM through kinetic mixing with the hypercharge gauge boson, giving rise to the so-called dark photon scenario~\cite{Holdom:1985ag,Pospelov:2007mp,Essig:2013lka}. 
Despite their rich phenomenological implications, such conventional constructions do not, in general, address the origin of neutrino masses.

In recent years, a number of frameworks has been proposed to simultaneously address neutrino masses and DM within a common symmetry structure~\cite{Ma:2006km,Mandal:2019oth,Huong:2026idm}. 
While these scenarios are theoretically appealing and phenomenologically rich as relating the issue of dark matter and that of neutrino masses, they often ad hoc involve extended particle contents with additional symmetries, hence the predictive possibility of the model is reduced. This leaves room for more economical frameworks, i.e. minimal realizations, in which both neutrino masses and DM emerge from a unified and minimal setup, enhancing the predictive level as well as probing the idea.

More recently, a class of models according to this direction has emerged in which dark gauge symmetries are employed for both neutrino mass generation and DM stability. In particular, the framework proposed in ~\cite{Dong:2025rhs} introduces a $U(1)_D$ gauge symmetry that acts nontrivially on right-handed neutrinos, leading to an anomaly-free structure and a residual discrete symmetry that stabilizes a DM candidate. In that construction, neutrino masses arise from a hybrid mechanism combining seesaw and radiative contributions, the so-called scotoseesaw. While this setup provides an elegant connection between neutrino mass generation and DM existence, the origin of the small lepton-number violating parameter is not dynamically explained and remains separated from the underlying symmetry structure.

In this work, we propose an alternative framework in which a local $U(1)_D$ gauge symmetry plays a central and unified role. First, it forbids Majorana mass terms for sterile neutrinos at tree level, thereby ensuring lepton number conservation in the limit of symmetry conservation. Second, it enables the radiative generation of the small lepton-number violating parameter $\mu$ through interactions within the dark sector. In particular, we construct a radiative inverse seesaw scenario in which $\mu$ arises at the one-loop level, providing a dynamical and symmetry-protected origin for its smallness.

After spontaneous symmetry breaking, the $U(1)_D$ symmetry is reduced to a residual $Z_2$ symmetry, under which a subset of fields is odd. This residual symmetry ensures the stability of the lightest dark sector particle, offering a viable DM candidate. Remarkably, the same interactions responsible for stabilizing DM also generate the small Majorana masses required for the inverse seesaw mechanism, thereby establishing a direct connection between neutrino mass generation and DM physics. The light neutrino masses in this framework are given by \begin{equation} m_\nu \simeq m_D M^{-1} \mu (M^T)^{-1} m_D^T, \end{equation} where the smallness of $m_\nu$ is governed by the loop-suppressed parameter $\mu$. The radiative origin of $\mu$ ensures that it vanishes in the limit of restored symmetry, rendering its smallness technically natural.
Unlike scenarios based on classical scale invariance or the so-called neutrino option~\cite{Brivio:2017dfq}, our approach does not aim to generate all mass scales dynamically. Instead, we focus on providing a minimal and robust explanation for the origin of neutrino masses and DM within a unified framework governed by a dark gauge symmetry. This leads to a predictive setup with rich phenomenological implications in neutrino physics, DM, and colliders.

The rest of this paper is organized as follows. We first present the model and its field content in Sec.~\ref{model}. We then diagonalize the scalar potential to identify scalar mass spectrum in \ref{chv}. The derivation of the neutrino mass matrix and the radiative origin of the $\mu$ parameter are given in Sec.~\ref{neutrino}. The interplay between non-unitarity of lepton mixing matrix and lepton flavor violation (LFV) is then discussed in Sec.~\ref{Unita}. Whist, the structure of the gauge sector is outlined in Sec.~\ref{gauge}. We subsequently examine the Higgs phenomenology in Sec.~\ref{Higgs} and analyze the dark matter phenomenology together with relevant experimental constraints in Sec.~\ref{DM}. Finally, our conclusions are presented in Sec.~\ref{conclusion}. Further technical details on the Casas--Ibarra parametrization in the inverse seesaw framework can be found in Appendix~\ref{app:CI_inverse}. A discussion of fermionic dark matter mixing and gauge interactions is provided in Appendix~\ref{app:fermionDM}.

\section{The model}
\label{model}
We consider an extension of the SM by introducing an additional dark gauge symmetry $U(1)_D$, under which the particle content is enlarged by new scalar and fermion fields, apart from the dark gauge boson. Only a subset of these fields transforms nontrivially under $U(1)_D$. The new fermions are singlets under the SM gauge group and are required to satisfy anomaly cancellation conditions. They are therefore introduced in chiral pairs with opposite $U(1)_D$ charges. In particular, three generations of chiral fermion pairs $\nu_{aR}$ and $N_{aR}$, for $a=1,2,3$, carrying charges $D=2$ and $D=-2$, respectively, are introduced as part of the inverse seesaw structure. The $U(1)_D$ gauge symmetry allows Dirac mass terms between $\nu_{aR}$ and $N_{aR}$, while forbidding Majorana mass terms for each field at tree level. As a result, the tree-level inverse seesaw contribution to active neutrino masses is absent.
In addition, two chiral fermion pairs $\chi_{nR}$ and $\chi'_{nR}$ for $n=1,2$ with charges $D=1$ and $D=-1$, respectively, are minimally introduced in order for viable neutrino phenomenology. Indeed, these fields participate in the radiative generations of Majorana masses for $N_{aR}$ and $\nu_{aR}$ after symmetry breaking, which require at least two different generated masses for each kind of right-handed neutrinos and also provide fermionic dark matter candidates.
Furthermore, the new scalar sector contains two electrically neutral singlet fields, $\varsigma$ and $\sigma$, and an additional scalar doublet $\eta$. As for the inverse seesaw scheme, $\eta$ necessarily couples $\nu_{aR}$ to usual lepton doublets, while $\varsigma$ ($\varsigma^*$) especially couples $\nu_{aR}$ ($N_{aR}$) to $\chi'_{nR}$ $(\chi_{nR})$. Lastly, $\sigma$ couples to $\chi_R\chi_R$ as well as $\chi'_R\chi'_R$. The charge assignments of all fields under $SU(3)_C \otimes SU(2)_L \otimes U(1)_Y \otimes U(1)_D$ are summarized in Table~\ref{tab1}.

\begin{table}[h]
	\bc
	\begin{tabular}{lccccc}
		\hline\hline 
		Field & $SU(3)_C$ & $SU(2)_L$ & $U(1)_Y$ & $U(1)_D$ & $P_D$ \\
		\hline
		$l_{aL} = \begin{pmatrix}
			\nu_{aL}\\
			e_{aL}
		\end{pmatrix}$ & 1 & 2 & $-1/2$ & $0$ & + \\
		$e_{aR}$ & 1 & 1 & $-1$ & $0$ & + \\
		$\nu_{aR}$ & 1 & 1 & $0$ & $2$ & + \\
		$N_{aR}$ & 1 & 1 & $0$ & $-2$ & +  \\
		$\chi_{nR}$ & 1 & 1 & $0$ & $+1$ & $-$\\
		$\chi_{nR}^\prime$ & 1 & 1 & $0$ & $-1$ & $-$ \\
		$q_{aL} = \begin{pmatrix}
			u_{aL}\\
			d_{aL}
		\end{pmatrix}$ & 3 & 2 & $1/6$ & $0$  & +\\
		$u_{aR}$ & 3 & 1 & $2/3$ & $0$ & +\\
		$d_{aR}$ & 3 & 1 & $-1/3$ & $0$ & + \\
		$H = \begin{pmatrix}
			H^+\\
			H^0
		\end{pmatrix}$ & 1 & 2 & $1/2$ & $0$ & +\\
		$\eta = \begin{pmatrix}
			\eta^0\\
			\eta^-
		\end{pmatrix}$ & 1 & 2 & $-1/2$ & $-2$ & + \\
		$\sigma$ & 1 & 1 & $0$ & $2$ & + \\
		$\varsigma$ & 1 & 1 & $0$ & $-1$ & $-$\\
		\hline\hline 
	\end{tabular}
	\caption{\label{tab1} Field content of the model.}
	\ec
\end{table}

The spontaneous breaking of $U(1)_D$ is driven by the vacuum expectation values (VEVs) of $\sigma$ and $\eta$. The VEV of $\sigma$ generates Majorana masses for $\chi_{nR}$ and $\chi'_{nR}$, while the VEV of $\eta$ induces Dirac neutrino masses. After symmetry breaking, the $U(1)_D$ gauge symmetry is broken down to a residual discrete $Z_2$ symmetry, $P_D=(-1)^D$. This residual symmetry transforms on a field as
\begin{equation}
	Z_2:\quad \phi \;\to\; (-1)^{D(\phi)}\,\phi,
\end{equation}
where $D(\phi)$ is the $D$-charge of $\phi$. It is noted that $P_D$ is easily derived from the breaking of $U(1)_D$ by a scalar field carrying an even charge, such as $\phi=\sigma,\eta^0$. Fields with odd (even) $U(1)_D$ charge are therefore $Z_2$-odd (even), hence they are collected in Tab. \ref{tab1}. This $Z_2$ symmetry ensures the stability of the lightest $Z_2$-odd particle. The scalar $\varsigma$ is odd under the residual $Z_2$ symmetry and does not acquire any VEV due to the $Z_2$ conservation. It provides a scalar dark matter candidate. Alternatively, the lightest of odd-fermions $\chi_{nR}, \chi'_{nR}$ provides a fermionic dark matter candidate. The odd sectors $\chi_{nR}$, $\chi'_{nR}$, and $\varsigma$ mediate the one-loop generations of Majorana mass terms for both $N_{aR}$ and $\nu_{aR}$, the so-called scotogenic generated masses. 

In addition to the Yukawa interactions of the SM, the Yukawa Lagrangian involving the new fermions is given by
\bea
-\mathcal{L}_Y &\supset& y^\nu_{ab} \bar{\l}_{aL} \eta \nu_{bR} + M_{ab} \bar{N}^c_{aR}\nu_{bR}+ \kappa_{an}^N \bar{N}^c_{aR} \chi_{nR}\varsigma^*+ \kappa_{an}^\nu \bar{\nu}^c_{aR} \chi_{nR}^\prime \varsigma \nonumber  \\ &+&m_{\chi \chi^\prime}\bar{\chi}_{nR}^c \chi_{mR}^\prime 
+y_{\chi_{R}}\bar{\chi}_{nR}^c\chi_{nR}\sigma^*
+y_{\chi_{R}}^\prime \bar{\chi}_{nR}^{\prime c}\chi_{nR}^\prime\sigma + H.c.
\eea The SM quark sector remains unchanged. Their masses and those of charged leptons are the same standard model. The given interactions would deliver appropriate neutrino masses when taking into account radiative corrections to  $N_{aR}$ and $\nu_{aR}$ Majorana masses, as well as suitable new-fermion masses, as all indicated in subsequent sections.

The scalar potential invariant under $SU(3)_C \otimes SU(2)_L \otimes U(1)_Y \otimes U(1)_D$ is given by
\begin{align}
	V &= \mu_1^2 H^\dagger H + \mu_2^2 \eta^\dagger \eta + \mu_3^2 \sigma^\dagger \sigma + \mu_4^2 \varsigma^\dagger \varsigma \nonumber \\
	&\quad + \lambda_1 (H^\dagger H)^2 + \lambda_2 (\eta^\dagger \eta)^2 + \lambda_3 (\sigma^\dagger \sigma)^2 + \lambda_4 (\varsigma^\dagger \varsigma)^2 \nonumber \\
	&\quad + (H^\dagger H)\left( \lambda_5 \eta^\dagger \eta + \lambda_6 \sigma^\dagger \sigma + \lambda_7 \varsigma^\dagger \varsigma \right) \nonumber \\
	&\quad + (\eta^\dagger \eta)\left( \lambda_8 \sigma^\dagger \sigma + \lambda_9 \varsigma^\dagger \varsigma \right) 
	+ \lambda_{10} (\sigma^\dagger \sigma)(\varsigma^\dagger \varsigma) + \lambda_{11} (H^\dagger \eta)(\eta^\dagger H) \nonumber \\
	&\quad + \left\{ 
	\lambda_{12} (\eta^T i\sigma_2 H)\,\varsigma^{\ast}\varsigma^{\ast} 
	+ \mu_{123} (\eta^T i\sigma_2 H)\,\sigma 
	+ \mu_{34} \sigma \varsigma \varsigma 
	+ \text{H.c.} 
	\right\}.
\end{align}
Assuming CP conservation, all scalar couplings are taken to be real. The necessary conditions for the potential to be bounded from below, as well as having expected vacuum structure, are $\mu^2_{1,2,3}<0$, $\mu^2_4>0$, and $\la_{1,2,3,4}>0$. That said, the neutral scalar components can be expanded as
\begin{align}
	H^0 &= \frac{1}{\sqrt{2}}(v + S_1 + i A_1), \qquad 
	\eta^0 = \frac{1}{\sqrt{2}}(u + S_2 + i A_2), \nonumber \\
	\sigma &= \frac{1}{\sqrt{2}}(\Lambda + S_3 + i A_3), \qquad 
	\varsigma = \frac{1}{\sqrt{2}}(S_4 + i A_4).
\end{align} As mentioned, the VEVs $u$ and $\Lambda$ break $U(1)_D$ down to the residual $Z_2$ symmetry, $P_D=(-1)^D$, under which the fields $\varsigma$, $\chi_{nR}$, and $\chi'_{nR}$ are odd, while all other fields are even. As a consequence of the $Z_2$ conservation, the mixing between $Z_2$-even and $Z_2$-odd states is forbidden. To be consistent with the standard model, we impose $\Lambda\gg u,v$.

\section{\label{chv} Scalar mass spectrum}

The VEVs of scalar fields obey the conditions of potential minimization, such as
\begin{align}
	\mu_1^2 + \lambda_1 v^2 + \frac{\mu_{123}}{\sqrt{2}} \frac{u \Lambda}{v} + \frac{\lambda_5}{2} u^2 + \frac{\lambda_6}{2} \Lambda^2 &= 0, \nonumber \\
	\mu_2^2 + \lambda_2 u^2 + \frac{\mu_{123}}{\sqrt{2}} \frac{v \Lambda}{u} + \frac{\lambda_5}{2} v^2 + \frac{\lambda_8}{2} \Lambda^2 &= 0, \nonumber \\
	\mu_3^2 + \lambda_3 \Lambda^2 + \frac{\mu_{123}}{\sqrt{2}} \frac{u v}{\Lambda} + \frac{\lambda_6}{2} v^2 + \frac{\lambda_8}{2} u^2 &= 0.
	\label{linear1}
\end{align} These conditions would shift relevant scalar fields to physical fields to be obtained. 

First, the charged scalar fields $H^\pm$ and $\eta^\pm$ mix, leading to the mass eigenstates $\mathcal{G}_{W^\pm}$ and $\mathcal{H}^\pm$, which are defined as
\begin{align}
	\mathcal{G}_{W^\pm} &= \cos\theta\, H^\pm - \sin\theta\, \eta^\pm, \nonumber \\
	\mathcal{H}^\pm &= \sin\theta\, H^\pm + \cos\theta\, \eta^\pm,
\end{align}
where $\tan\theta = \frac{u}{v}$. The state $\mathcal{G}_{W^\pm}$ is massless and is identified as the Goldstone boson absorbed by the $W^\pm$ gauge bosons, while $\mathcal{H}^\pm$ corresponds to a physical charged scalar with mass, given by 
\begin{equation}
	m^2_{\mathcal{H}^\pm} = \frac{1}{\sin 2\theta} \left( \lambda_{11} u v - \sqrt{2}\,\mu_{123} \Lambda \right).
\end{equation}

The CP-odd scalar fields mix through the following mass terms,
\begin{align}
	V \supset \frac{1}{2}
	\begin{pmatrix}
		A_1 & A_2 & A_3
	\end{pmatrix}
	\mathcal{M}_A^2
	\begin{pmatrix}
		A_1 \\ A_2 \\ A_3
	\end{pmatrix},
\end{align}
where the mass matrix is given by
\begin{align}
	\mathcal{M}_A^2 = -\frac{\mu_{123}}{\sqrt{2}}
	\begin{pmatrix}
		\Lambda \tan\theta & \Lambda & u \\
		\Lambda & \Lambda \cot\theta & v \\
		u & v & v \tan\theta'
	\end{pmatrix},
\end{align}
with $\tan\theta' = \frac{u}{\Lambda}$.  Diagonalizing the mass matrix, one obtains two massless states, $\mathcal{G}_Z$ and $\mathcal{G}_{Z'}$, which are identified as the Goldstone bosons absorbed by the neutral gauge bosons $Z$ and $Z'$, respectively, as well as one massive CP-odd scalar $\mathcal{A}$. The corresponding eigenstates are given by
\begin{align}
	\mathcal{G}_Z &= \frac{1}{\sqrt{1 - \sin^2\theta \sin^2\theta'}} 
	\left[
	- A_1 \cos\theta 
	+ A_2 \sin\theta \cos^2\theta' 
	+ A_3 \sin\theta \sin\theta' \cos\theta'
	\right], \nonumber \\
	\mathcal{G}_{Z'} &= A_2 \sin\theta' + A_3 \cos\theta', \nonumber \\
	\mathcal{A} &= -\frac{1}{\sqrt{1 - \sin^2\theta \sin^2\theta'}} 
	\left[
	A_1 \sin\theta \cos\theta' 
	+ A_2 \cos\theta \cos\theta' 
	+ A_3 \cos\theta \sin\theta'
	\right].
\end{align}
The mass of the $\mathcal{A}$ is given as
\bea
 m^2_{\mathcal{A}}=-\frac{\mu_{123}}{\sqrt{2}uv\La} \left( v^2 \La^2+u^2 \La^2+u^2v^2\right).
\eea

The CP-even scalar fields $S_1, S_2, S_3$ mix, and their mass terms can be written as
\begin{align}
	V \supset \frac{1}{2}
	\begin{pmatrix}
		S_1 & S_2 & S_3
	\end{pmatrix}
	\mathcal{M}_S^2
	\begin{pmatrix}
		S_1 \\ S_2 \\ S_3
	\end{pmatrix},
\end{align}
where the mass matrix is given by
\begin{align}
	\mathcal{M}_S^2 =
	\begin{pmatrix}
		2\lambda_1 v^2 - \frac{\mu_{123}}{\sqrt{2}} \frac{u\Lambda}{v} &
		\lambda_5 u v + \frac{\mu_{123}}{\sqrt{2}} \Lambda &
		\lambda_6 v \Lambda + \frac{\mu_{123}}{\sqrt{2}} u \\[6pt]
		\lambda_5 u v + \frac{\mu_{123}}{\sqrt{2}} \Lambda &
		2\lambda_2 u^2 - \frac{\mu_{123}}{\sqrt{2}} \frac{v\Lambda}{u} &
		\lambda_8 u \Lambda + \frac{\mu_{123}}{\sqrt{2}} v \\[6pt]
		\lambda_6 v \Lambda + \frac{\mu_{123}}{\sqrt{2}} u &
		\lambda_8 u \Lambda + \frac{\mu_{123}}{\sqrt{2}} v &
		2\lambda_3 \Lambda^2 - \frac{\mu_{123}}{\sqrt{2}} \frac{uv}{\Lambda}
	\end{pmatrix}.
	\label{CP1}
\end{align}
The minimization conditions in Eqs.~(\ref{linear1}) indicate that the parameters $\mu_{123}$ and $\Lambda$ can be of the same order. In this regime, the CP-even mass matrix approximately yields the following physical eigenstates:
\begin{align}
	h &\simeq \frac{1}{\sqrt{1+\kappa_h^2}}
	\left( S_1 \cos\theta + S_2 \sin\theta + S_3 \kappa_h \right), \nonumber \\
	\mathcal{H} &\simeq \frac{1}{\sqrt{1+\kappa_{\mathcal{H}}^2}}
	\left( -S_1 \sin\theta + S_2 \cos\theta + S_3 \kappa_{\mathcal{H}} \right), \nonumber \\
	\mathcal{H}' &\simeq \frac{1}{\sqrt{1 + (\kappa_h')^2 + (\kappa_{\mathcal{H}}')^2}}
	\left( S_1 \kappa_h' + S_2 \kappa_{\mathcal{H}}' + S_3 \right),
\end{align}
where $\kappa_h, \kappa_{\mathcal{H}}, \kappa_h', \kappa_{\mathcal{H}}' \sim \mathcal{O}\!\left(\frac{v,u}{\Lambda,\,\mu_{123}}\right)$ represent next-to-leading-order corrections. Their masses are approximately given by
\begin{align}
	m_h^2 &\simeq \frac{2}{u^2 + v^2}
	\left( \lambda_1 v^4 + \lambda_2 u^4 + \lambda_5 u^2 v^2 \right)
	+ \mathcal{O}\!\left(\frac{v,u}{\Lambda,\,\mu_{123}}\right), \nonumber \\
	m_{\mathcal{H}}^2 &\simeq -\frac{\sqrt{2}}{\sin 2\theta}\,\mu_{123}\Lambda
	+ 2(\lambda_1 + \lambda_2 - \lambda_5)\frac{u^2 v^2}{u^2 + v^2}, \nonumber \\
	m_{\mathcal{H}'}^2 &\simeq 2\lambda_3 \Lambda^2 - \frac{\mu_{123}}{\sqrt{2}\Lambda} uv.
\end{align} The state $h$ is identified as the SM-like Higgs boson, while the others $\mathcal{H}$ and $\mathcal{H}'$ are new neutral Higgs fields. 

Due to the presence of the $\mu_{34}$ and $\lambda_{12}$ terms, the real and imaginary components of the $Z_2$-odd scalar $\varsigma$ are physical fields by themselves, acquiring different masses,
\begin{align}
	m^2_{\Re \varsigma} &= \frac{1}{2}
	\left( 2\mu_4^2 + \lambda_9 u^2 + \lambda_7 v^2 + \lambda_{10} \Lambda^2
	+ 2\sqrt{2}\,\mu_{34}\Lambda + 2\lambda_{12} u v \right), \nonumber \\
	m^2_{\Im \varsigma} &= \frac{1}{2}
	\left( 2\mu_4^2 + \lambda_9 u^2 + \lambda_7 v^2 + \lambda_{10} \Lambda^2
	- 2\sqrt{2}\,\mu_{34}\Lambda - 2\lambda_{12} u v \right).
\end{align}The mass splitting is therefore given by
\begin{equation}
	m^2_{\Re \varsigma} - m^2_{\Im \varsigma}
	= 2\sqrt{2}\,\mu_{34}\Lambda + 2\lambda_{12} u v,
	\label{split1}
\end{equation}
which plays a crucial role in radiative generation of Majorana masses for $N_{aR}$ and $\nu_{aR}$.

\section{Neutrino mass generation}
\label{neutrino}
\begin{figure}[H]
	\bc
	\includegraphics[scale=0.8]{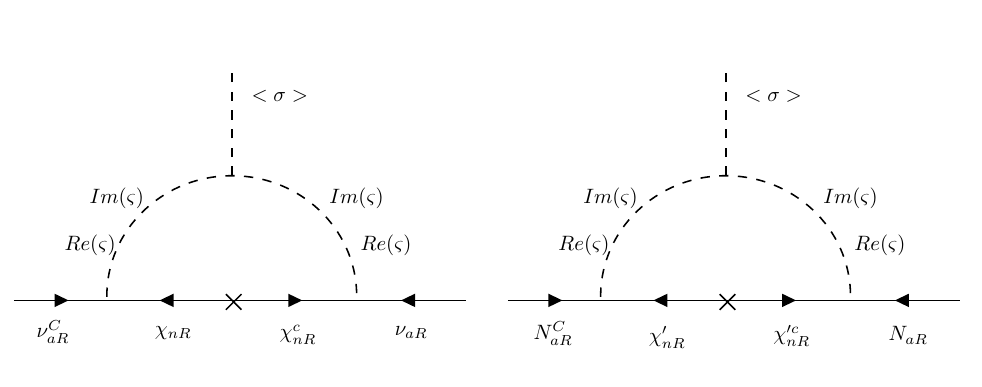}
	\caption[]{\label{Fig1} One-loop Feynman diagram contributing to the Majorana masses, $\mu, \mu^\prime$.}
	\ec
\end{figure}

In this section, we derive the radiatively generated Majorana masses and discuss their implications for the inverse seesaw mechanism.
We now turn to the generation of neutrino masses in this model. In the basis
\(
(\nu_L,\ \nu_R^c,\ N_R^c)
\),
the neutral lepton mass matrix takes the form
\begin{equation}
	M_\nu=
	\begin{pmatrix}
		0 & m_D & 0\\
		m_D^T & \mu' & M\\
		0 & M^T & \mu
	\end{pmatrix},
\end{equation}
where the submatrices \(M\) and
\(
m_D=\frac{y^\nu}{\sqrt2}u
\)
are generated at tree level, while \(\mu\) and \(\mu'\) arise radiatively at one loop through the diagrams shown in Fig.~\ref{Fig1}. 

Explicitly, one finds
\begin{align}
	\mu_{ab}
	&=
	\sum_{n=1}^2
	\frac{\kappa^N_{an}\kappa^N_{bn}\,m_{\chi_n}}{16\pi^2}
	\left[
	\frac{m^2_{\Re\varsigma}}{m^2_{\Re\varsigma}-m^2_{\chi_n}}
	\ln\frac{m^2_{\Re\varsigma}}{m^2_{\chi_n}}
	-
	\frac{m^2_{\Im\varsigma}}{m^2_{\Im\varsigma}-m^2_{\chi_n}}
	\ln\frac{m^2_{\Im\varsigma}}{m^2_{\chi_n}}
	\right],
	\nonumber\\
	\mu'_{ab}
	&=
	\sum_{n=1}^2
	\frac{\kappa^\nu_{an}\kappa^\nu_{bn}\,m_{\chi'_n}}{16\pi^2}
	\left[
	\frac{m^2_{\Re\varsigma}}{m^2_{\Re\varsigma}-m^2_{\chi'_n}}
	\ln\frac{m^2_{\Re\varsigma}}{m^2_{\chi'_n}}
	-
	\frac{m^2_{\Im\varsigma}}{m^2_{\Im\varsigma}-m^2_{\chi'_n}}
	\ln\frac{m^2_{\Im\varsigma}}{m^2_{\chi'_n}}
	\right].
	\label{mu}
\end{align}
It is clear from Eq.~\eqref{mu} that in the limit
\(
m^2_{\Re\varsigma}=m^2_{\Im\varsigma}
\),
the contributions of the CP-even and CP-odd scalar components cancel exactly, so that \(\mu=\mu'=0\). Therefore, non-zero Majorana masses are generated only because of the mass splitting between \(\Re\varsigma\) and \(\Im\varsigma\), as given in Eq.~(\ref{split1}).  Assuming a small splitting,
\begin{equation}
	\Delta m_\varsigma^2
	\equiv
	m^2_{\Re\varsigma}-m^2_{\Im\varsigma}
	=
	2\sqrt{2}\mu_{34}\Lambda+2\lambda_{12}uv
	\ll
	m^2_{\Re\varsigma}+m^2_{\Im\varsigma}
	\equiv 2m_\varsigma^2,
\end{equation}
the loop functions can be expanded to first order in \(\Delta m_\varsigma^2\). One then obtains
\begin{align}
	\mu_{ab}
	&\simeq
	\sum_{n=1}^2
	\frac{\kappa^N_{an}\kappa^N_{bn}}{16\pi^2}
	\,\Delta m_\varsigma^2\,m_{\chi_n}
	\left[
	\frac{1}{m_\varsigma^2-m_{\chi_n}^2}
	+
	\frac{m_{\chi_n}^2\ln\!\left(m_{\chi_n}^2/m_\varsigma^2\right)}
	{\left(m_\varsigma^2-m_{\chi_n}^2\right)^2}
	\right],
	\nonumber\\
	\mu'_{ab}
	&\simeq
	\sum_{n=1}^2
	\frac{\kappa^\nu_{an}\kappa^\nu_{bn}}{16\pi^2}
	\,\Delta m_\varsigma^2\,m_{\chi'_n}
	\left[
	\frac{1}{m_\varsigma^2-m_{\chi'_n}^2}
	+
	\frac{m_{\chi'_n}^2\ln\!\left(m_{\chi'_n}^2/m_\varsigma^2\right)}
	{\left(m_\varsigma^2-m_{\chi'_n}^2\right)^2}
	\right].
	\label{mu_approx}
\end{align}

Hence, both \(\mu\) and \(\mu'\) are suppressed not only by the loop factor \(1/(16\pi^2)\), but also by the small scalar mass splitting \(\Delta m_\varsigma^2\), which itself is controlled by the parameters \(\mu_{34}\) and \(\lambda_{12}\). In the limit
\(
\kappa^{N,\nu},\,\mu_{34},\,\lambda_{12}\to 0
\),
the Lagrangian exhibits an enhanced \(U(1)_\varsigma\) symmetry under which only \(\varsigma\) is charged. In this limit, the contributions from the CP-even and CP-odd scalar components cancel exactly, leading to vanishing \(\mu\) and \(\mu'\). According to ’t Hooft naturalness, the smallness of the parameters \(\kappa^{N,\nu}\), \(\mu_{34}\), and \(\lambda_{12}\), as well as the induced Majorana masses, is therefore technically natural, being protected by the approximate symmetry.

In the inverse seesaw regime,
\begin{equation}
	\mu,\mu' \ll m_D \ll M,
\end{equation}
the effective light neutrino mass matrix is approximately given by
\begin{equation}
	\mathcal M_\nu^{\rm light}
	\simeq
	m_D\,M^{-T}\,\mu\,M^{-1}\,m_D^T
	+\mathcal O(\mu\mu',\,\mu^2,\,{\mu'}^2).
	\label{mnu_eff}
\end{equation}
Defining the active--sterile mixing matrix as
\begin{equation}
	\Theta \simeq m_D M^{-T},
\end{equation}
Eq.~\eqref{mnu_eff} can be rewritten as
\begin{equation}
	\mathcal M_\nu^{\rm light}\simeq \Theta\,\mu\,\Theta^T.
\end{equation}
This relation makes the origin of small neutrino masses particularly transparent: they are simultaneously controlled by the small lepton-number violating parameter \(\mu\) and by the active--sterile mixing \(\Theta\). The active--sterile mixing is itself restricted, providing an additional suppression mechanism for light neutrino masses. 

A more quantitative analysis of these bounds, together with complementary constraints from charged lepton flavor violation, will be presented in the following sections.
We note in passing that the radiatively induced \(\mu\) term in the present model is reminiscent of the mechanism discussed in Ref.~\cite{Abada:2021yot}, where the lepton-number violating parameter is also tied to dark-sector dynamics. In our setup, however, the resulting DM phenomenology is qualitatively different, as will be elaborated in the next section.

In summary, light neutrino masses arise from the interplay of loop suppression, a small lepton-number violating scale, and constrained active--sterile mixing. This structure renders the inverse seesaw framework both phenomenologically viable and theoretically well motivated.

\section{Interplay between non-unitary and flavor violation}
\label{Unita}

\subsection{Non-unitarity}
Non-unitarity effects arise from the mixing between active and sterile neutrinos, which directly induces a deviation of the effective leptonic mixing matrix from exact unitarity. After block diagonalization of the full neutral fermion mass matrix, the charged-current interaction is governed by an effective mixing matrix $N$, which can be expressed as follows~\cite{Antusch:2006vwa,Grimus:2000vj}
\begin{equation}
	N \simeq \left(1 - \frac{1}{2}\Theta\Theta^\dagger\right) U_\nu,
	\label{eq:N_nonunitary}
\end{equation}
where $U_\nu$ is the unitary matrix that diagonalizes the light neutrino mass matrix, and $\Theta$ parametrizes the mixing between the active neutrinos and the heavy sterile states.
In the inverse seesaw limit in question, $\mu \ll m_D \ll M$, the active--sterile mixing matrix is approximately given by
\begin{equation}
	\Theta \simeq m_D^\dagger M^{-1}.
	\label{eq:Theta_def}
\end{equation}
Accordingly, the deviation from unitarity can be quantified by the Hermitian matrix
\begin{equation}
	\eta \equiv \mathbf{1} - N N^\dagger,
\end{equation}
which, at leading order in $\Theta$, reduces to~\cite{Antusch:2006vwa,Grimus:2000vj}
\begin{equation}
	\eta \simeq \frac{1}{2}\Theta\Theta^\dagger.
	\label{eq:eta_def}
\end{equation}

Precision electroweak data and flavor observables impose stringent bounds on non-unitarity, typically requiring $\eta_{ii} \lesssim \mathcal{O}(10^{-3})$ and $|\eta_{ij}| \lesssim \mathcal{O}(10^{-5})$. These constraints directly restrict the size of the active--sterile mixing $\Theta$, independently of the specific origin of neutrino masses.
In the inverse seesaw framework, the light neutrino mass matrix is approximately given by
\begin{equation}
	m_\nu \simeq \Theta \, \mu \, \Theta^T.
	\label{eq:mnu_inverse_seesaw}
\end{equation}
Therefore, for fixed light neutrino masses, a smaller value of the lepton-number violating parameter $\mu$ requires a larger active--sterile mixing $\Theta$. As a result, non-unitarity constraints prevent $\mu$ from becoming arbitrarily small.  Since the same combination $\Theta\Theta^\dagger$ also controls charged lepton flavor violating observables, non-unitarity bounds have direct implications for LFV phenomenology. This establishes a close interplay between the two effects and makes non-unitarity constraints a powerful and complementary probe of the inverse seesaw parameter space.

\subsection{Lepton flavor violation}
We now turn to examine the one-loop radiative decays $l_i \to l_j \gamma$, whose branching ratios are given by~\cite{Ilakovac:1994kj,Deppisch:2004fa}
\begin{eqnarray}
	\mathrm{BR}\left(l_i \to l_j \gamma \right)
	&=&
	\frac{\alpha_W^3 s_W^2 m_{l_i}^5}{256\pi^2 m_W^4 \Gamma_i}
	\left| G_{ij} \right|^2,
	\label{eq:BR_lilgamm}
	\\
	G_{ij}
	&\simeq&
	\sum_{k=1}^{3}
	N^*_{ik} N_{jk}
	\, G_\gamma\!\left(\frac{m_{\nu_k}^2}{m_W^2}\right)
	\nonumber\\
	&&
	+\; 2 \sum_{n}
	\Theta^*_{in}\Theta_{jn}
	\, G_\gamma\!\left(\frac{M_n^2}{m_W^2}\right),
	\label{eq:Gij_LFV}
	\\
	G_\gamma(x)
	&=&
	\frac{10 - 43x + 78x^2 - 49x^3 + 18x^3 \ln x + 4x^4}
	{12(1-x)^4},
	\label{eq:Ggamma_loop}
\end{eqnarray}
where $\Gamma_i$ denotes the total decay width of the charged lepton $l_i$.  The amplitude receives contributions from both light and heavy neutrino states propagating in the loop. 

In typical inverse seesaw scenarios, the light neutrino contribution is strongly suppressed by the tiny neutrino masses, whereas the dominant contribution arises from the heavy sterile states through the active--sterile mixing $\Theta$.
This structure makes the connection with non-unitarity manifest. Indeed, both the non-unitarity parameter $\eta$ and the LFV amplitudes are governed by the same mixing combination $\Theta\Theta^\dagger$. Consequently, any constraint that suppresses non-unitarity effects also tends to reduce the predicted rates for radiative charged lepton decays. Conversely, regions of parameter space with sizable LFV signals generally correspond to enhanced active--sterile mixing and therefore to larger deviations from unitarity.

\subsection{Numerical result and discussion}
In this section, we perform a numerical analysis of the interplay between neutrino mass generation, non-unitarity effects, and charged lepton flavor violation in the radiative inverse seesaw framework. We carry out a random scan of the model parameter space by varying the entries of the lepton-number violating Majorana mass matrix $\mu$ within the following range:
\begin{equation}
	10^{-2}~\mathrm{keV} \leq \mu_i \leq 10^{2}~\mathrm{keV},
\end{equation}
while the heavy neutrino masses are taken within:
\begin{equation}
	100~\mathrm{GeV} \leq M_i \leq 5~\mathrm{TeV}.
\end{equation}

The light neutrino sector is fixed to be consistent with current oscillation data. We assume normal mass ordering and take the light neutrino masses as
\begin{equation}
	m_1 = 10^{-3}~\mathrm{eV}, \quad 
	m_2 = \sqrt{m_1^2 + \Delta m_{21}^2}, \quad 
	m_3 = \sqrt{m_1^2 + \Delta m_{31}^2}.
\end{equation}
We adopt the current best-fit values of neutrino oscillation parameters from the global analysis of Ref.~\cite{Esteban:2020cvm}, such as 
\bea
\Delta m_{21}^2 = 7.42 \times 10^{-5}~\mathrm{eV}^2, \quad
\Delta m_{31}^2 = 2.517 \times 10^{-3}~\mathrm{eV}^2.
\eea
The leptonic mixing matrix $U_{\rm PMNS}$ is constructed using the best-fit values of the mixing angles and CP phase. 

The Dirac mass matrix is constructed using the Casas--Ibarra parametrization. In this framework, the active--sterile mixing matrix is given in Appendix  \ref{app:CI_inverse} by
\begin{equation}
	\Theta = U_{\rm PMNS} \, \sqrt{m_\nu^{\rm diag}} \, R \, \mu^{-1/2},
\end{equation}
where $R$ is a complex orthogonal matrix satisfying $R^T R = \text{ I} $. In our numerical analysis, $R$ is fully parameterized in terms of three complex angles, allowing for a general exploration of the parameter space. The deviation from unitarity of the leptonic mixing matrix is quantified by
\begin{equation}
	\eta = \frac{1}{2} \Theta \Theta^\dagger.
\end{equation}

To identify phenomenologically viable regions of the parameter space, we impose the most stringent experimental upper bound on the branching ratio of the lepton flavor violating decay $\mu^+ \to e^+ \gamma$, as reported by the MEG collaboration at 90\% confidence level~\cite{MEG:2016leq}, as follows
\begin{equation}
	\mathrm{BR}(\mu^+ \to e^+ \gamma) < 4.2 \times 10^{-13}.
\end{equation}
\begin{figure}[H]
	\centering
	\includegraphics[width=0.6\textwidth]{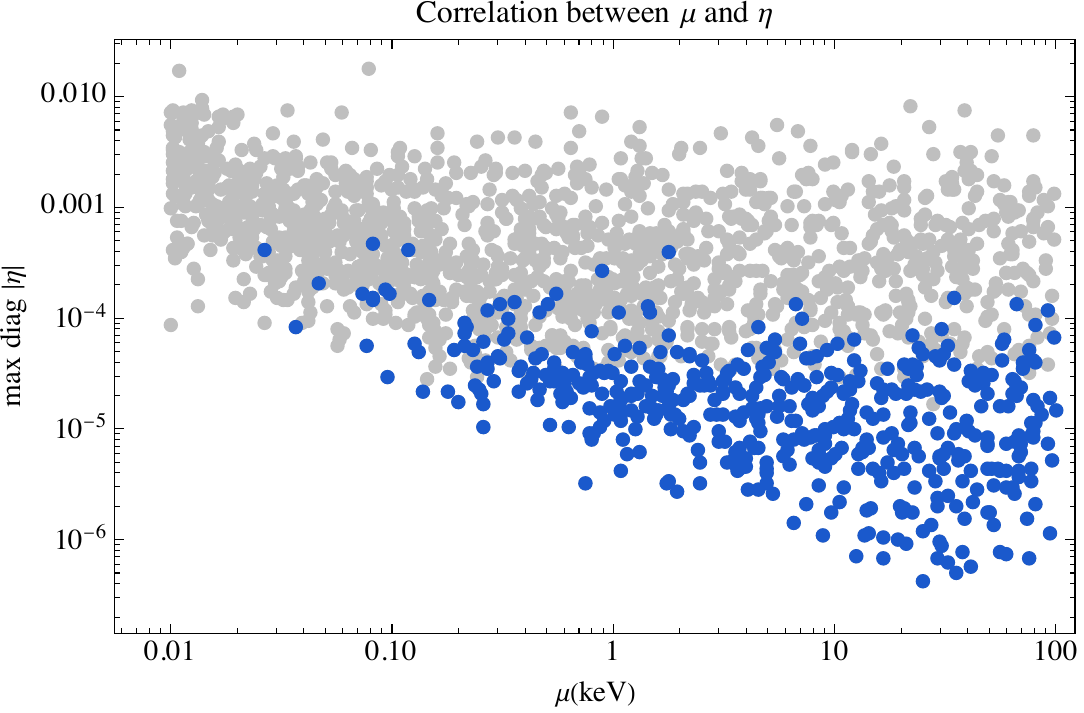}
	\caption{
		Correlation between the lepton-number-violating parameter $\mu$ and the size of non-unitarity effects, quantified by $\max |\eta_{ii}|$, in the radiative inverse seesaw framework. }
	\label{fig:mu_eta}
\end{figure}

Figure~\ref{fig:mu_eta} illustrates the correlation between the lepton-number violating parameter $\mu$ and the size of non-unitarity effects, characterized by $\max |\eta_{ii}|$. The gray points represent the full parameter scan, while the blue points satisfy the experimental bound on $\mathrm{BR}(\mu \to e \gamma)$. It is important to note that the scatter plots shown in this work are not statistical distributions. The density of points depends on the adopted scanning procedure, including the choice of parameter ranges and the sampling method, and therefore does not carry direct physical significance. The physical information is instead encoded in the overall structure of the parameter space and in the correlations among observables. In particular, the plots illustrate the regions that are compatible with experimental constraints, as well as the existence of correlations between the model parameters and low-energy observables.  The observed correlation between the lepton-number violating parameter $\mu$ and 
the non-unitarity measure $\eta$ can be understood from the structure of the active--sterile mixing matrix. In the present framework,
the magnitude of $\Theta$ increases as $\mu$ decreases. It follows that $\eta$ scales approximately as $\mu^{-1}$. Therefore, smaller values of $\mu$ are associated with larger active--sterile mixing and correspondingly larger deviations from unitarity. This behavior explains the trend observed in the numerical results, where points with smaller $\mu$ tend to yield larger values of $\eta$, while larger $\mu$ suppresses these effects.
\begin{figure}[H]
	I	\centering
	\includegraphics[width=0.6\textwidth]{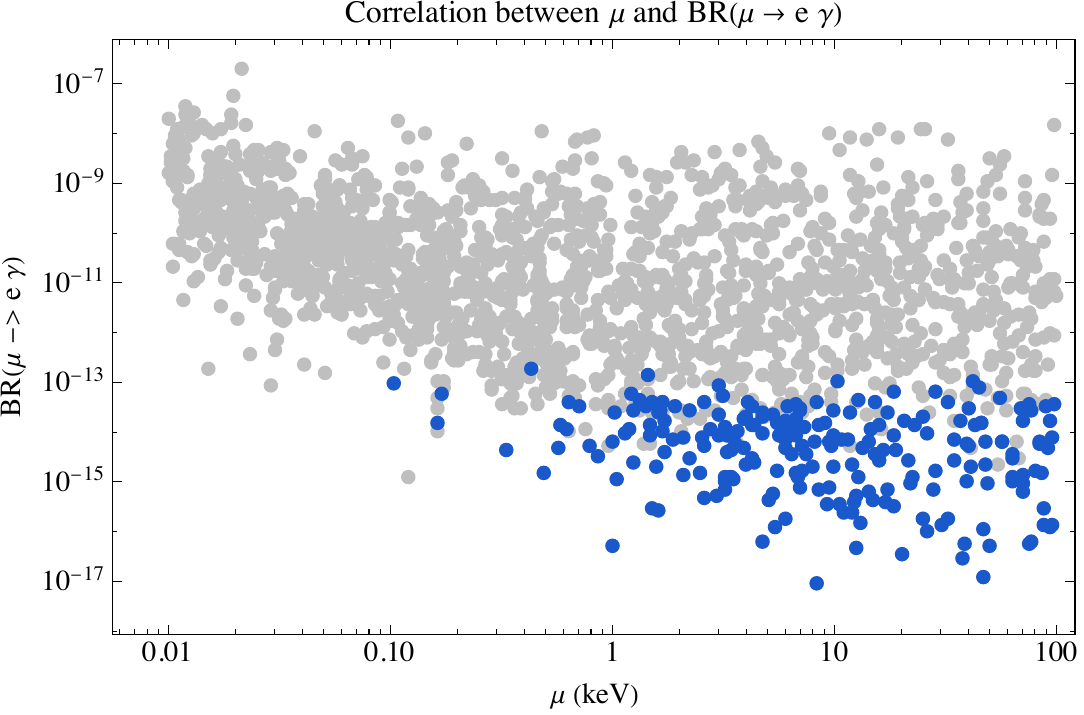}
	\caption{
		Correlation between the radiatively generated lepton-number-violating parameter $\mu$ and the size of non-unitarity effects, quantified by $\max |\eta_{ii}|$, in the inverse seesaw framework with a full complex Casas--Ibarra parametrization. 
	}
	\label{fig:mu_BR}
\end{figure}

Figure~\ref{fig:mu_BR} shows the correlation between the lepton-number violating parameter $\mu$ and the branching ratio $\mathrm{BR}(\mu \to e \gamma)$. The gray points correspond to the full parameter scan, while the blue points satisfy experimental constraints on non-unitarity~\cite{Antusch:2006vwa}.
A general trend of enhanced LFV for decreasing $\mu$ can be observed, which follows from the inverse seesaw relation,
\begin{equation}
	m_\nu \simeq \Theta \, \mu \, \Theta^T,
\end{equation}
implying that smaller values of $\mu$ require larger active--sterile mixing $\Theta$. Since LFV amplitudes are governed by $\Theta \Theta^\dagger$, this leads to an overall increase in $\mathrm{BR}(\mu \to e \gamma)$ at small $\mu$.
However, a significant spread of several orders of magnitude is present at fixed $\mu$. This reflects the dependence of LFV observables on the underlying flavor structure, including complex phases and possible cancellations among different heavy neutrino contributions.  Imposing non-unitarity constraints strongly suppresses the allowed mixing, resulting in the blue points populating the lower region of the plot and significantly reducing the predicted LFV rates, well below the current experimental bound from MEG~\cite{MEG:2016leq}.
This demonstrates that non-unitarity constraints provide a more stringent restriction on the parameter space than the current LFV bounds in a large region of the model.

\section{Gauge boson masses and mixing}
\label{gauge}
Gauge bosons acquire masses after the spontaneous breaking of gauge symmetry
\[
SU(3)_C \otimes SU(2)_L \otimes U(1)_Y \otimes U(1)_D 
\;\longrightarrow\;
SU(3)_C \otimes U(1)_Q \otimes Z_2.
\]
The masses arise from the scalar kinetic terms
\begin{equation}
	\mathcal{L}_{\text{kin}}^{\text{scalar}} 
	= \sum_{S=H, \eta, \sigma} 
	\left( D^\mu S \right)^\dag \left( D_\mu S \right),
\end{equation}
where
\begin{equation}
	D_\mu = \partial_\mu 
	+ ig_s T_n G_{n\mu} 
	+ ig t_a A_{a\mu}
	+ ig^\prime Y B_\mu 
	+ ig^{\prime\prime} D C_\mu.
\end{equation}

The charged gauge bosons acquire masses
\begin{equation}
	m_W^2 = \frac{g^2}{4}(u^2 + v^2).
\end{equation}
In the neutral sector, the fields \((A_{3\mu}, B_\mu, C_\mu)\) mix after symmetry breaking, by contrast. In general, both mass mixing and kinetic mixing effects can be present.

\subsection*{A. Without kinetic mixing}
We first consider the case where kinetic mixing is absent. In this case, the neutral gauge bosons arise solely from mass mixing induced by the scalar vacuum expectation values, i.e. symmetry breaking. The electroweak rotation is defined as
\begin{align}
	A_\mu &= s_W A_{3\mu} + c_W B_\mu, \nonumber \\
	Z_\mu^0 &= c_W A_{3\mu} - s_W B_\mu,
\end{align}
with $\tan\theta_W = g'/g$. The photon $A_\mu$ remains massless, while the massive neutral states arise from the mixing between $Z_\mu^0$ and $C_\mu$,
\begin{equation}
	\begin{pmatrix}
		Z_\mu \\
		Z'_\mu
	\end{pmatrix}
	=
	\begin{pmatrix}
		\cos\alpha & \sin\alpha \\
		-\sin\alpha & \cos\alpha
	\end{pmatrix}
	\begin{pmatrix}
		Z_\mu^0 \\
		C_\mu
	\end{pmatrix}.
\end{equation}
The mixing angle $\alpha$ is determined by
\begin{equation}
	\tan 2\alpha =
	\frac{
		-\dfrac{2 g g_D}{c_W}u^2
	}{
		4g_D^{\,2}(u^2+\Lambda^2) - \dfrac{g^2}{4c_W^2}(u^2+v^2)
	},
\end{equation} which is small due to $u,v\ll \La$. 
The corresponding masses are
\begin{align}
	m_Z^2 &= \frac{g^2}{4c_W^2}(u^2+v^2)\cos^2\alpha
	- \frac{g g_D}{c_W}u^2 \sin 2\alpha
	+ 4g_D^{\,2}(u^2+\Lambda^2)\sin^2\alpha, \nonumber \\
	m_{Z'}^2 &= \frac{g^2}{4c_W^2}(u^2+v^2)\sin^2\alpha
	+ \frac{g g_D}{c_W}u^2 \sin 2\alpha
	+ 4g_D^{\,2}(u^2+\Lambda^2)\cos^2\alpha.
\end{align}
In the limit $\alpha \to 0$, or $u,v \ll \Lambda$, one recovers the SM expression for $m_Z$, while $Z'$ acquires a mass at the $U(1)_D$ breaking scale.

Since SM fermions are neutral under $U(1)_D$, the gauge boson $C_\mu$ does not couple directly to SM currents. The neutral current interaction is therefore inherited entirely from the $Z_\mu^0$ component,
\begin{equation}
	\mathcal L_{\rm int}
	\supset
	\frac{g}{c_W} Z_\mu^0 \sum_f \bar f \gamma^\mu
	\left( T_3^f - s_W^2 Q_f \right) f.
\end{equation}
After rotating to the mass eigenstates, the couplings of the physical gauge bosons can be expressed in terms of those of the SM neutral current. Defining
\begin{equation}
	g_L^{Z_0,f} = \frac{g}{c_W}(T_3^f - s_W^2 Q_f), 
	\qquad
	g_R^{Z_0,f} = -\frac{g}{c_W}s_W^2 Q_f,
\end{equation}
one finds
\begin{align}
	g_{L,R}^{Z,f} &= \cos\alpha\, g_{L,R}^{Z_0,f}, \\
	g_{L,R}^{Z',f} &= -\sin\alpha\, g_{L,R}^{Z_0,f}.
\end{align}
This implies that, in the absence of kinetic mixing, the interaction of $Z'$ with the SM is entirely controlled by the mixing angle $\alpha$.

\subsection*{B. Including kinetic mixing}
We now include kinetic mixing between \(U(1)_Y\) and \(U(1)_D\),
\begin{equation}
	\mathcal L_{\rm kin}
	\supset
	-\frac14 B_{\mu\nu} B^{\mu\nu}
	-\frac14 C_{\mu\nu} C^{\mu\nu}
	-\frac{\epsilon}{2} B_{\mu\nu} C^{\mu\nu}.
\end{equation}
For \(\epsilon \ll 1\), the kinetic terms can be diagonalized and canonically normalized via a field redefinition. After electroweak symmetry breaking, the photon remains massless, while the massive neutral sector can be described in the \((Z_\mu^0,\, C_\mu)\) basis. To leading order in \(\epsilon\), the neutral gauge boson mass matrix is given by
\begin{equation}
	M^2 \simeq
	\begin{pmatrix}
		m_{Z_0}^2 & \Delta^2 \\
		\Delta^2 & m_{CC}^2
	\end{pmatrix},
\end{equation}
This can be expressed in terms of the parameters
\begin{align}
	m_{Z_0}^2 &= \frac{g^2}{4c_W^2}(u^2+v^2), \quad
	\Delta^2 =
	-\frac{g g_D}{c_W}u^2
	+\frac{\epsilon g^2 s_W}{4c_W^2}(u^2+v^2), \\
	m_{CC}^2 &=
	4 g_D^{\,2}(u^2+\Lambda^2)
	-\frac{2\epsilon g g_D s_W}{c_W}u^2.
\end{align}
The physical masses are obtained as
\begin{equation}
	m_{Z,Z'}^2=
	\frac12\left[
	m_{Z_0}^2+m_{CC}^2
	\mp
	\sqrt{(m_{CC}^2-m_{Z_0}^2)^2+4(\Delta^2)^2}
	\right],
\end{equation}
and the mixing angle \(\alpha\) is determined by
\begin{equation}
	\tan 2\alpha =
	\frac{2\Delta^2}{m_{CC}^2 - m_{Z_0}^2}.
\end{equation}
Thus, the mixing angle receives contributions from both Higgs-induced mass mixing and the kinetic-mixing parameter \(\epsilon\).

At leading order, the kinetic mixing can be removed by the field redefinition
\begin{equation}
	B_\mu \;\to\; B_\mu + \epsilon\, C_\mu,
\end{equation}
which induces an interaction between \(C_\mu\) and the hypercharge current,
\begin{equation}
	\mathcal L_{\rm int}
	\supset
	\epsilon g' C_\mu \sum_f Y_f \bar f \gamma^\mu f.
\end{equation}
After diagonalization of the mass matrix, the couplings of the physical states can be expressed as follows
\begin{align}
	g_{L,R}^{Z,f} &= \cos\alpha\, g_{L,R}^{Z_0,f}
	+ \sin\alpha\, \epsilon g' Y_{L,R}^f , \\
	g_{L,R}^{Z',f} &= -\sin\alpha\, g_{L,R}^{Z_0,f}
	+ \cos\alpha\, \epsilon g' Y_{L,R}^f .
\end{align}

In the limit $\alpha \ll 1$ and $\epsilon \ll 1$, the SM $Z$ couplings receive only subleading corrections, whereas the $Z'$ interactions involve both mass-mixing and kinetic-mixing contributions at leading order. The model thus features two distinct portals connecting the dark sector to the SM: a mass-mixing portal controlled by $\alpha$, and a kinetic-mixing portal controlled by $\epsilon$. The corresponding induced couplings are typically suppressed, ensuring consistency with LEP bounds. Their interplay governs the phenomenology of dark matter and collider observables.

\subsection*{C. Electroweak constraints from \(Z\)–\(Z'\) mixing}

An important consequence of the \(Z\)–\(Z'\) mixing is its impact on electroweak precision observables, most notably on the tree-level rho parameter,
\begin{equation}
	\rho \equiv \frac{m_W^2}{m_Z^2 c_W^2}.
\end{equation}
In the SM, one has \(\rho=1\) at tree level. In the present model, however, the mixing in the neutral gauge boson sector shifts the physical \(Z\)-boson mass and induces a deviation in \(\rho\).
In the small-mixing regime, one finds
\begin{equation}
	\Delta\rho \equiv \rho-1
	\simeq
	\frac{(\Delta^2)^2}{m_{Z_0}^2\left(m_{CC}^2-m_{Z_0}^2\right)}.
	\label{eq:Delta_rho_general}
\end{equation}
This expression shows that electroweak precision data strongly constrain off-diagonal mixing term \(\Delta^2\), hence both the Higgs-induced and kinetic-mixing contributions.

To explore the viable parameter space, we perform a numerical scan over
\begin{align}
	u &\in [1,\,245]~\text{GeV}, \quad \Lambda \in [0.5,\,5]~\text{TeV}, \\
	g_D &\in [10^{-3},\,1], \quad \epsilon \in [-10^{-2},\,10^{-2}].
\end{align}
Here, \(u\) denotes the vacuum expectation value associated with the breaking of \(U(1)_D\), satisfying \(u^2 + v^2 = v_{\rm EW}^2\).
The global electroweak fit yields~\cite{PDG2022,deBlas:2020ofp}
\begin{equation}
	\Delta\rho = (3.9 \pm 1.9)\times 10^{-4}.
\end{equation}
Requiring compatibility within $2\sigma$ of the experimental value, we adopt the conservative bound, such as $\Delta\rho < 2.7 \times 10^{-3}$.

\begin{figure}[t]
	\centering
	\includegraphics[width=0.68\textwidth]{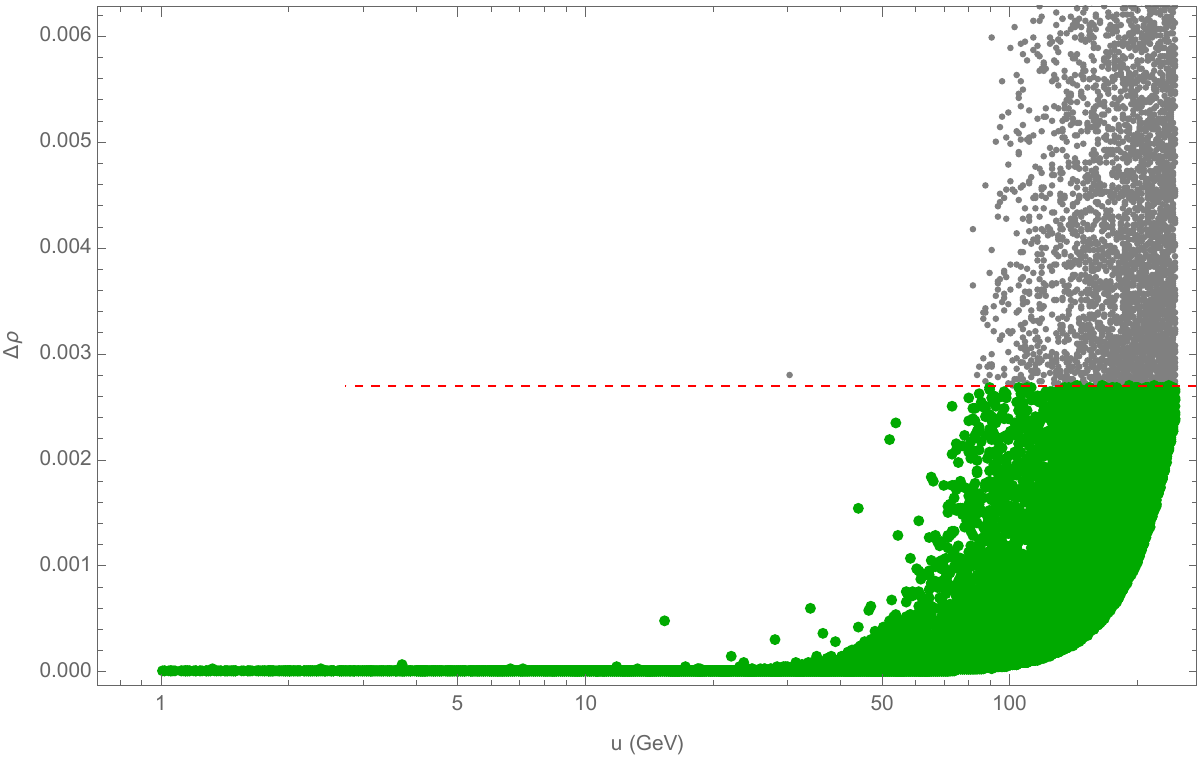}
	\caption{
		Distribution of the electroweak precision parameter $\Delta\rho$ as a function of the vacuum expectation value $u$. The gray points denote all physical points in the parameter scan, while the green points satisfy the electroweak precision constraint $\Delta\rho < 2.7\times 10^{-3}$. The red dashed line indicates the corresponding upper bound.
	}
	\label{fig:delta_rho_u}
\end{figure}
Figure~\ref{fig:delta_rho_u} shows $\Delta\rho$ as a function of $u$. 
As the plot represents a projection onto the $(u,\Delta\rho)$ plane, different combinations of $\Lambda$, $g_D$, and $\epsilon$ can correspond to the same value of $u$, leading to a spread in $\Delta\rho$ at fixed $u$. 
The figure demonstrates that the electroweak precision observable $\Delta\rho$ imposes a strong constraint on the symmetry-breaking scale $u$, with large values of $u$ allowed only in restricted regions of parameter space associated with suppressed mixing or finely tuned cancellations. This behavior can be understood from the structure of the neutral gauge boson mixing parameter,
\begin{equation}
	\Delta^2 =
	-\frac{g g_D}{c_W}u^2
	+\frac{\epsilon g^2 s_W}{4c_W^2} v_{\rm EW}^2,
\end{equation}
which receives contributions from both Higgs-induced mass mixing and kinetic mixing. 
For small $u$, the term proportional to $\epsilon v_{\rm EW}^2$ can be relevant, whereas for larger $u$ the $u^2$ term dominates. 
Consequently, the allowed parameter space is controlled by the interplay of these two contributions rather than by $u$ alone.

In particular, parameter points with relatively large $u$ can still satisfy the electroweak bound if a cancellation occurs in $\Delta^2$, thereby suppressing $\Delta\rho$. 
Such viable points correspond to finely tuned regions of parameter space rather than generic configurations. To illustrate this effect, we consider the benchmark point
\begin{equation}
	\Lambda = 1~\mathrm{TeV}, \quad u = 80~\mathrm{GeV}, \quad g_D = 0.3.
\end{equation}
For this choice, $\Delta\rho$ can be approximated as
\begin{equation}
	\begin{aligned}
		\Delta\rho
		&\simeq
		6.9\times 10^{-4}
		- 3.9\times 10^{-3}\,\epsilon
		+ 5.5\times 10^{-3}\,\epsilon^2
		+ \mathcal{O}(\epsilon^3).
	\end{aligned}
\end{equation}
It is clear that in the small-$\epsilon$ regime, the dependence on $\epsilon$ is mild. 
However, its contribution can induce cancellations in $\Delta^2$, thereby suppressing $\Delta\rho$ and enlarging the viable parameter space.

 \section{Higgs boson phenomenology}
\label{Higgs}
In this work, we employ the $\kappa$-framework \cite{LHCHiggsCrossSectionWorkingGroup:2013rie, LHCHiggsCrossSectionWorkingGroup:2012nn} to connect the properties
of the Higgs boson in  the considered model to experimental  measurements of the Higgs boson production and decay modes. The modifier parameters, $\{\kappa_i \}$, with $i=l,u,d,W,Z$ defined as the ratios of the Higgs boson couplings to particles $i$ to their corresponding SM values, are given as 
\bea
\kappa_{l_a}&=&\kappa_{u_a}=\kappa_{d_a}=\kappa_W= 2-\frac{1}{\sqrt{1+\kappa_h^2}}\equiv \kappa_f, \label{ka_f}\eea  \bea
\kappa_Z &\simeq& c_{\al}+\frac{12g_D}{g}c_W s^2_{\theta}s_{\al}. 
\eea
Here $\ka_Z$ is obtained in the limit keeping only the first order of $Z-Z'$ mixing angle $\al$, i.e $s_{\al}^2\to 0$. For loop-level coupling of $ggh, \ga \ga h$ and $Z\ga h$, the modifier parameters can be expressed as a function of the SM coupling modifier \cite{ParticleDataGroup:2024cfk}
\bea
\kappa^2_g&=&0.99952 k_f^2,  \crn
\kappa_\ga^2&=& 1.59 \kappa_W^2-0.66 \kappa_W \kappa_t+0.07 \kappa_t^2, \crn
\ka_{Z\ga}^2&=&1.12\ka_W^2-0.15\ka_W\ka_f+0.03\ka_f^2. 
\eea
It is important to note that here we have skipped the charged Higgs  $\mathcal{H}^{\pm}$ contribution to above parameters. This is explained by the reason that this kind of contribution is suppressed by factor $m_W^2/m_{\mathcal{H}^{\pm}}^2\sim \mathcal{O}(10^{-4}-10^{-3})$ with $m_{\mathcal{H}^{\pm}}$ is at TeV scale. Thus, this charged Higgs contribution to parameters $\ka_{\ga,Z\ga}$ is significantly smaller compared to other ones and can be ignored. 

\begin{table}[H]
	\bc
	\begin{tabular}{lcccc}
		\hline\hline 
		Observables & ATLAS Run 2+Run   & CMS Run 2 \\
		\hline
		$\ka_{g}$ & $0.99^{+0.07}_{-0.06}$  & $0.91^{+0.07}_{-0.06}$ \\
		$\ka_{\ga}$ & $0.97\pm0.06$ & $1.10\pm 0.07$ \\
		$\ka_{Z}$ & $0.96\pm 0.05 $ & $1.07\pm 0.06$  \\
		$\ka_{W}$ & $1.00\pm 0.05 $ & $1.03\pm 0.06$ \\
		$\ka_{b}$ & $0.89 ^{+0.10}_{-0.09} $ & $0.98^{+0.13}_{-0.12}$   \\
		$\ka_{t}$ & $0.99\pm 0.09$ & $0.92^{+0.09}_{-0.08}$   \\
		$\ka_{\mu}$ & $1.04^{+0.23}_{-0.30}$ & $1.09^{+0.20}_{-0.22}$   \\
		$\ka_{\tau}$ & $0.93\pm 0.07$ & $0.92\pm0.08$   \\
		\hline\hline 
	\end{tabular}
	\caption[]{\label{expHiggs} The second and third column show the $1\sigma$ experimental constraints by ATLAS and CMS for modified couplings $\ka_i$.}
	\ec
\end{table}  
The predicted observables will be compared with latest ATLAS Run 2+Run 3 and CMS Run 2 results \cite{ParticleDataGroup:2024cfk}, which are numerically listed in Table \ref{expHiggs}.
Firstly, let us give an estimation about some observables. The modified couplings given in Eq. (\ref{ka_f}) can be approximately expressed as $\ka_f\simeq 1+\ka_h^2/2$ since the suppressed coefficient $\ka_h\sim \mathcal{O}(u,v/\La)\sim \mathcal{O}(10^{-2}-10^{-3})\ll 1$. Therefore, these parameters remarkably close to unity $\ka_f \simeq 1$ and satisfy $1\sigma$ constraint for $\ka_{f},\ka_W$ reported by both ATLAS Run 2 and CMS Run 2, as shown in Table \ref{expHiggs}. Besides, enhanced by $\ka_f\simeq 1$, several observables depending on them such as $\ka_g$, $\ka_{\ga}$ are approximately $\simeq 1$, which meet the measurements of ATLAS. However, the predicted $\ka_{\ga}$ is smaller than its corresponding CMS constraint since $\ka_{\ga}^{\text{CMS}}\geq 1.04$. The predicted $\ka_{\ga}$ is satisfied CMS result if $\ka_{h}\geq  \sqrt{0.08}\simeq 0.28$, which conflicts with its desire value $\ka_h\sim \mathcal{O}(10^{-2}-10^{-3})$. 

On the other hand, the NP contribution affects strongly to remaining observable  $\ka_Z$. This observable depends on electroweak VEV $u$, new physics scale $\La$ and $U(1)_D$ coupling $g_D$ which will be randomly seeded as the following ranges $u\in[1,246]$ GeV, $\La\in[1,5]$ TeV, $g_D\in[10^{-3},4\pi]$ for numerical study. Additionally, the mixing angle $\al$ between SM $Z$ and new neutral gauge boson $Z'$ is suppressed $|\al|\leq 10^{-3}$ due to electroweak precision test \cite{ParticleDataGroup:2024cfk}. 

\begin{figure}[H]
	\bc
	\includegraphics[scale=0.30]{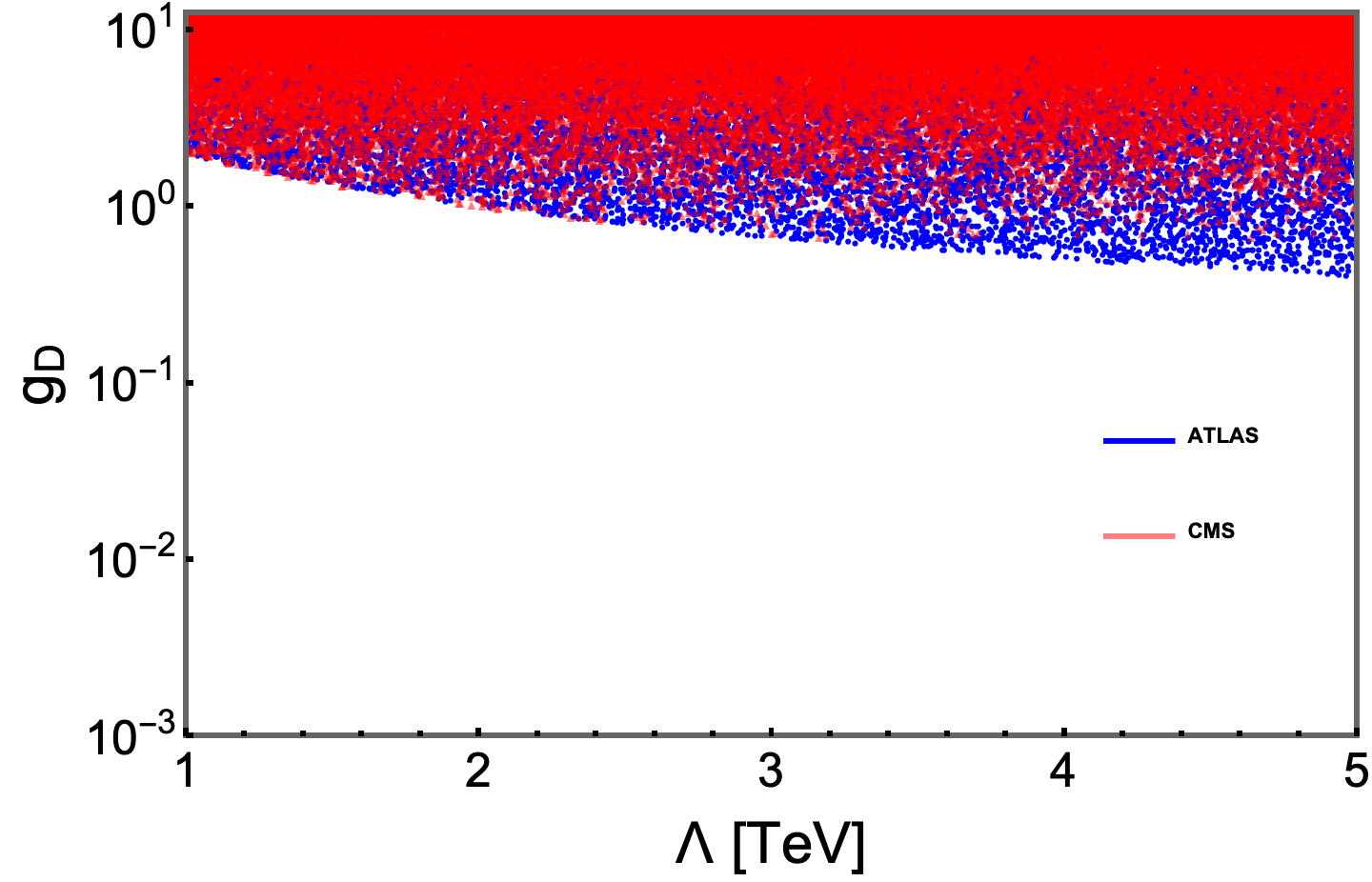}
	\includegraphics[scale=0.30]{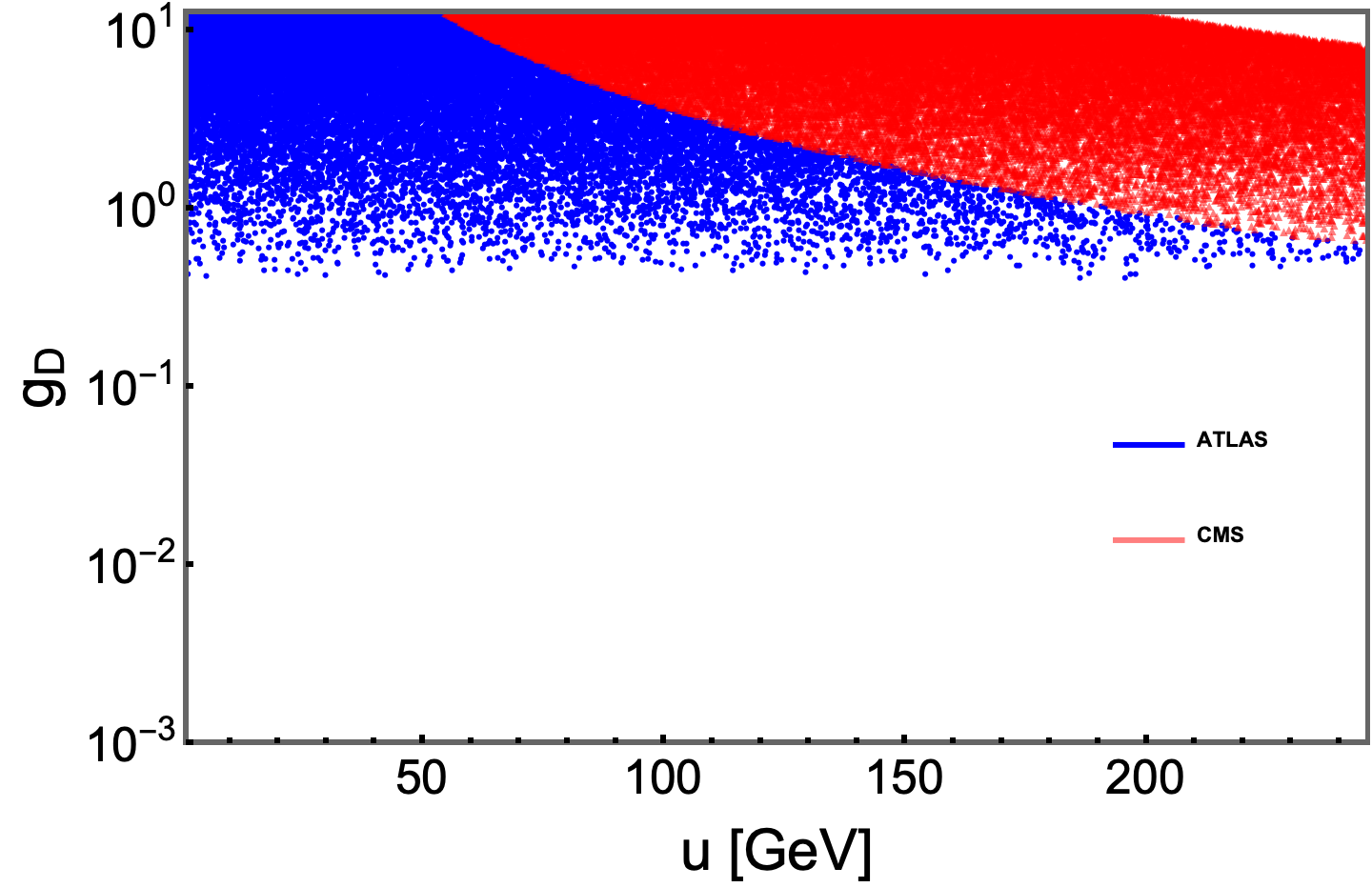}
	\caption[]{\label{ATLAS} The left and right panels show the correlation between $U(1)_D$ coupling $g_D$ with NP scale $\La$ and  electroweak scale $u$ satisfying ATLAS (blue), CMS (red) constraints of $\ka_Z$ given in Table \ref{expHiggs}, respectively.}
	\ec
\end{figure}
In the panels of Fig. \ref{ATLAS}, we show the relationship between coupling $g_D$ with VEV $\La$ and mixing angle $\theta$ satisfying the ATLAS, CMS constraints of $\ka_{Z}$ \cite{ATLAS:2022vkf}. We see that in the left panel, the values of $g_D$ tend to decrease when $\La$ raises. For instance, we get the lower bound limit of $g_D$ namely $g_D\geq 0.40$ for ATLAS and $g_D\geq 0.66$ for CMS constraints  at $\La=5$ TeV, and $g_D$ can attain the perturbation limit $4\pi$ at $\La=1$ TeV. For the right panel, we see that the constraint for $(u,g_D)$ are more sensitive for ATLAS and HL-LHC limits, compared to $(\La,g_D)$ in the left panel. The CMS limit gives stronger bound for electroweak VEV $u$ compared to ATLAS limit, i.e. we get $u\geq 55$ GeV for CMS while the whole range $u$ can fulfill the ATLAS result. In the case of maximum value $g_{D}=4\pi$, we get the upper bound of $u$ as $u\leq 56$ GeV and $u\leq 203$ GeV for ATLAS and CMS constraints, respectively.  
 
\section{Dark matter phenomenology}
\label{DM}
Within this framework, the stability of the DM candidate originates from the gauge structure of the model. In particular, the spontaneous breaking of the local \(U(1)_D\) symmetry leaves a residual discrete \(Z_2\) symmetry, under which a subset of dark-sector fields is odd, while all SM fields remain even. As a result, the lightest \(Z_2\)-odd state is stable and constitutes a viable DM candidate, provided it is electrically neutral. Notably, this stabilizing symmetry is not imposed by hand, but emerges as a remnant of the gauge symmetry.

Depending on the parameter space, the DM candidate can be realized either as a scalar state, associated with the components of \(\varsigma\), or as a fermionic state from the \(\chi_R\) sector. These possibilities correspond to distinct phenomenological regimes, as the dominant annihilation channels and detection prospects depend sensitively on the spin and interaction structure of the DM particle. In the following, we focus on two representative scenarios, corresponding to scalar and fermionic DM, respectively.

The relic abundance of DM is governed by a combination of scalar and gauge interactions. Scalar-mediated annihilation processes are controlled by Higgs mixing and portal couplings, while gauge-mediated channels proceed via exchange of the dark gauge boson \( Z'\) and even \(Z\) boson due to the mixing of two massive neutral gauge bosons. As a result, the observed relic density can be achieved in different regions of parameter space, including Higgs-portal-dominated, gauge-portal-dominated, or mixed scenarios. In addition, resonant annihilation and coannihilation effects may become relevant, depending on the mass spectrum of the dark sector.

A salient feature of the model is that the dark sector simultaneously accounts for dark matter (DM) and generates the radiative lepton-number-violating parameter $\mu$ entering the inverse seesaw mechanism. As a result, part of the interaction structure responsible for neutrino mass generation is directly linked to the dark sector. In particular, the couplings inducing $\mu$ also mediate interactions between the DM candidate and the sterile neutrino states $\nu_R$ and $N_R$.

However, this connection does not imply a one-to-one correspondence between the DM and neutrino sectors. The interactions responsible for the loop-induced generation of $\mu$ primarily involve DM couplings to sterile neutrinos and therefore contribute only indirectly to the processes governing the relic abundance. In typical realizations, the dominant annihilation channels are instead controlled by gauge interactions or scalar portal couplings, which do not enter the neutrino mass matrix. Moreover, the impact of the $\mu$-inducing interactions on DM observables is suppressed by the heavy mass scale of the sterile states and, where relevant, by the small active-sterile mixing.

Consequently, the parameter space does not reduce to a fully correlated structure, but retains a degree of independence between neutrino mass generation and DM dynamics. In the following, we focus on the DM phenomenology and do not pursue these correlations further. For concreteness, we analyze separately two representative realizations of the DM sector, corresponding to scalar and fermionic DM.

Let us present some input parameter before performing numerical scan. In order to pass the bounds from Higgs phenomenology, electroweak constraints, we adopt the following benchmark point $-\mu_{123}=\La=5$ TeV, $g_D=0.7$, $\ep=10^{-2}$, $u=238$ GeV and $v\simeq 60$ GeV.  For self-couplings $\la_{1,2,3,4}$, we set them at the same order, i.e. $\la_{1,2,3,4}=0.1$. With this setup, we can estimate the new particle masses as follow : $m_{H_1}\simeq 8500$ GeV, $m_{H_2}\simeq 2240$ GeV, $m_{H^{\pm}}\simeq m_{\mathcal{A}}\simeq 8500$ GeV and $m_{Z'}\simeq 7000$ GeV.  It is worth to note that the mass of new Higgs $H_1$ is always heavier than $H_2$. This is interpreted as $m_{H_1}^2\simeq \sqrt{2}\La^2/s_{2\theta}\geq \sqrt{2}\La^2$ (since $s_{2\theta}\leq 1$), thus  $m_{H_1}^2>m_{H_2}^2\simeq 2\la_3\La^2$ for $\la_3\sim \mathcal{O}(0.1)$. The quartic couplings $\la_{5}\simeq \la_{11},\la_6,\la_8$ are irrelevant and do affect to DM phenomenology, thus we skip these ones. As a result, the model leaves dimensionless couplings $\la_7,\la_9,\la_{10},\la_{12}$, mass dimension coupling $\mu_{34}$  as well as Yukawa couplings $\ka^{n,N},y^{(')}_{\chi}$ play important role to DM study. For numerical study, we scan these parameters in either small range $[10^{-5},10^{-3}]$ or larger $[10^{-3},10^{-1}]$. Here, the mass dimension coupling $\mu_{34}$ is assumed to be small since it relates to the suppressed condition $\Delta m_{\varsigma}^2=2\sqrt{2}\mu_{34}\La+2\la_{12}uv\ll 2m_{\varsigma}^2$, hence we consider $\mu_{34}\sim \mathcal{O}(10^{-3}-10^{-1})$ GeV. Besides, the masses of odd $Z_2$ particles are given in the range $[1,10]$ TeV. Here, we want to remark that mass of odd fermions $\chi^{(')}_{i}$ and Yukawa couplings $\ka^{N,\nu}$ are relatively constrained via the smallness of LVN parameters $\mu^{(')}_{i}$.  We summarize our choices of input parameters in the Table. \ref{inputDM}. In this work, the public package micrOMEGAs \cite{Alguero:2023zol} is used to calculate relic density, direct searches for DM. 

\begin{table}[H]
	\centering
	\begin{tabular}{|c|c|}
		\hline
		\text{Parameters} & \text{Ranges} \\ \hline
		$|\la_{i}|$ ($i=7,9,10,12$) & $[10^{-3},10^{-2}]$ \text{or} $[10^{-4},10^{-3}]$ \text{or fixed}    \\ \hline
		$|\mu_{34}|$ & $[10^{-3},10^{-1}]$ GeV \\ \hline
		$\ka^{n,N},y^{(')}_{\chi}$ & $[10^{-3},10^{-1}]$ \\ \hline
		$m_{\Re_{\varsigma},\Im_{\varsigma}}$ & $[0.01,10]$ TeV \\ \hline
		$m_{\chi^{(')}_i}$ & $[1,10]$ TeV \\ \hline
		$\La$ & $5$ TeV \\ \hline
	  $(v,u)$ & $(238,62.)$ GeV \\ \hline
		$\ep$ & $10^{-2}$ \\ \hline
		$g_D$ & $0.7$ \\ \hline
	\end{tabular}
	\caption{Input parameters ranges for numerically study of DM}
	\label{inputDM}
\end{table}
\subsubsection{Scalar DM}

A first possibility is that the lightest $Z_2$-odd neutral state is a scalar, identified with either $\Re \varsigma$ or $\Im \varsigma$. In this case, the DM phenomenology is primarily governed by the scalar sector, with the dominant annihilation channels controlled by Higgs portal interactions and, depending on the gauge structure, by additional gauge-mediated processes involving the dark gauge bosons $Z, Z'$.

\begin{figure}[H]
	\bc
	\includegraphics[scale=0.4]{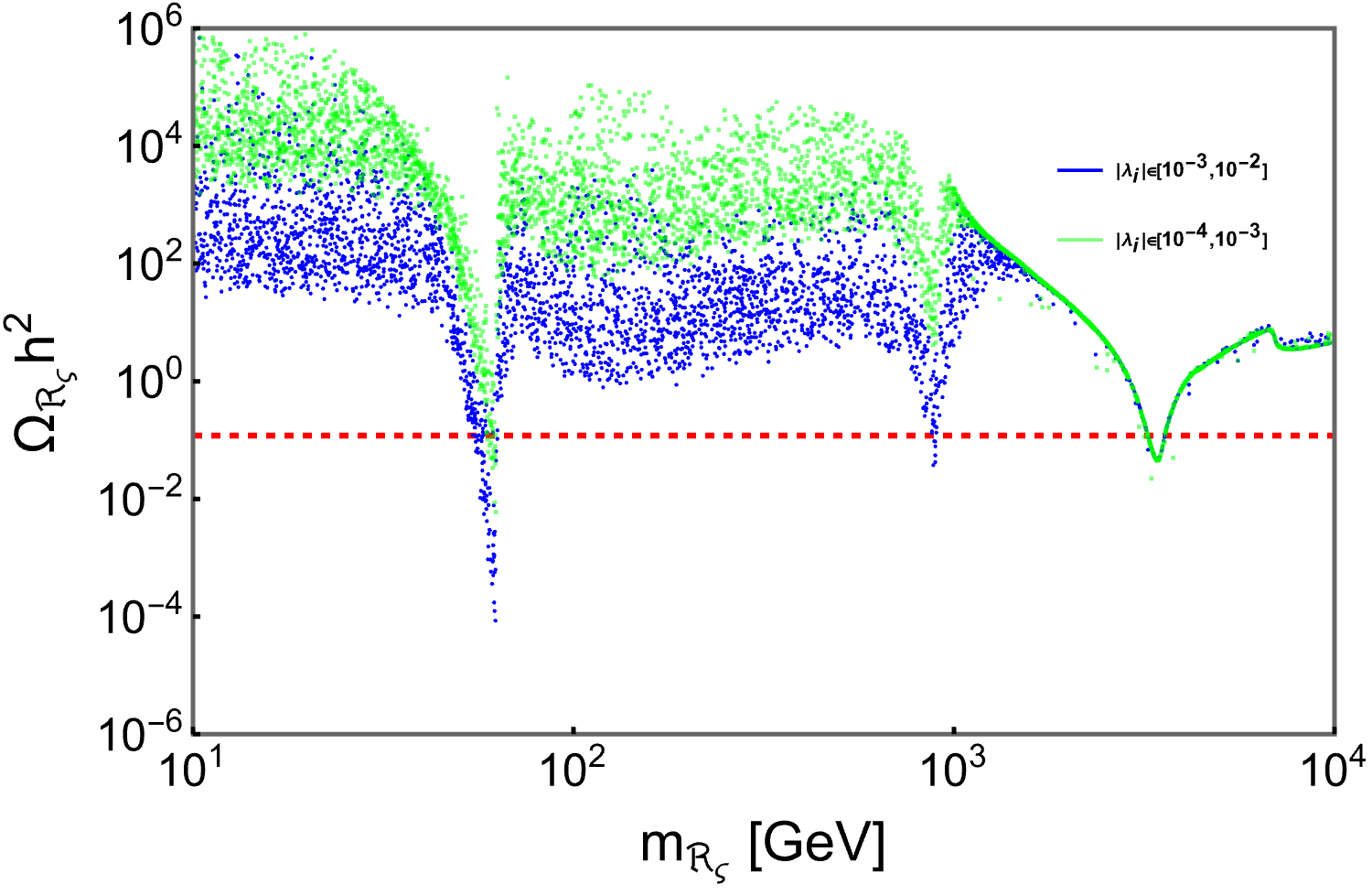}
	\caption[]{\label{scalarDM} The dependence of relic density $\Omega_{\Re_{\varsigma}}h^2$ as the function of its mass $m_{\Re_{\varsigma}}$. The dashed red line present the current experimental constraint of relic density  $\Omega h^2\simeq 0.12$ \cite{Planck:2018vyg}. }
	\ec
\end{figure}

In this work, we consider the real component $\Re_{\varsigma}$ is DM candidate, which is ensured by imposing condition $\Delta m_{\varsigma}^2<0$.  In Fig .\ref{scalarDM}, we show the dependence of relic density of $\Re_{\varsigma}$ as the function of its mass $m_{\Re_{\varsigma}}$ in the two difference range of quartic couplings $|\la_{i}|\in [10^{-3},10^{-2}]$ (blue points) and $|\la_i|\in[10^{-4},10^{-3}]$  (green points). The Yukawa couplings $\ka^{\nu,N}$ and odd $Z_2$ fermion particles are ensured to satisfy constraints from LFV decays and non-unitary effects. The figure indicates that the relic density is predicted to be lower in the scenario scalar couplings $|\la_{i}|\in [10^{-3},10^{-2}]$ than ones with $|\la_{i}|\in [10^{-4},10^{-3}]$. This can be understood due to the large quartic couplings $\la_i$ result in large thermally averaged cross-section times relative velocity $<\sigma v>_{\Re_{\varsigma},\Re_{\varsigma}
	\to X^*X}\sim|\la_i|^2$, where $X$ denotes for final states including either SM particles  $h,W,Z,f$ or new particles $H_1,H_2, H^{+},Z',\mathcal{A}$ if kinematic allowed. As a result, the relic density $\Omega h^2\sim 0.1/<\sigma v>_{\Re_{\varsigma},\Re_{\varsigma}
	\to X^*X}$ will be decreased.  In addition, the Fig. \ref{scalarDM} shows the existence of several resonances, in which the relic densities drop sharply and can obtain correct relic abundance $\Omega h^2\simeq 0.12$ \cite{Planck:2018vyg}.  The annihilations of $R_{\varsigma}$ are induced dominantly via the s-channel though Higgs $h,H_1,H_2$ and gauge portals $Z'$. When DM mass $m_{\Re_{\varsigma}}\simeq m_{\mathcal{P}}/2$ with $\mathcal{P}=\{h,H_1,H_2,Z'\}$, for instance $m_{\Re_{\varsigma}}\simeq m_h/2=62.5$ GeV, $m_{\Re_{\varsigma}}\simeq m_{H_2}/2\simeq 1120$ GeV and $m_{\Re_{\varsigma}}\simeq m_{Z'}/2\simeq 3500$ GeV, the thermally averaged cross-section can be written in a Breit-Wigner form 
\bea 
<\sigma v>_{\Re_{\varsigma},\Re_{\varsigma}
\to X,X}\simeq \fr{|\la_{\Re_{\varsigma}\Re_{\varsigma}\mathcal{P}}|^2|\Ga_{\mathcal{P}^*\to X^*X}(s)}{(s-m_{\mathcal{P}}^2)^2+m_{\mathcal{P}}^2[\Ga^{\text{tot}}_{\mathcal{P}}(s)]^2}  \label{DM-resonance}
\eea 
where $s\simeq 4m_{\Re_{\varsigma}}^2$, $\la_{\Re_{\varsigma}\Re_{\varsigma}\mathcal{P}}$ are vertexes depending on quartic couplings $\la_i$,  $\Ga_{\mathcal{P}^{*}\to X^*X}(s)$ is decay width of virtual $\mathcal{P}^* $, $\Ga^{\text{tot}}_{\mathcal{P}}(s)=\Ga_{\mathcal{P}\to X^*X}(s)+\Ga_{\mathcal{P}\to \Re_{\varsigma}\Re_{\varsigma}}(s)$ is the total decay width of $\mathcal{P}$. Therefore, as DM mass reaches the resonance points $s\simeq m_{\mathcal{P}}^2$, $<\sigma v>_{\Re_{\varsigma},\Re_{\varsigma}
	\to X^*X} \to  \fr{|\la_{\Re_{\varsigma}\Re_{\varsigma}\mathcal{P}}|^2|\Ga_{\mathcal{P}\to \text{SM,SM}}(s)}{[\Ga^{\text{tot}}_{\mathcal{P}}]^2}  $ tends to maximum value, thus making relic density $\Omega_{\Re_{\varsigma}} h^2\simeq 0.1/<\sigma v>_{\Re_{\varsigma},\Re_{\varsigma}
\to X^*X}\ll 1$. 
It is worth to mention that the scalar DM candidate $\Re_{\varsigma}$ satisfies the experimental result $\Omega h^2\simeq 0.12$ \cite{Planck:2018vyg} if its mass is around the resonance regions.  Moreover, if the couplings $\la_i$ are chosen to be larger $\la_i \geq 10^{-2}$, the relic density will be raised and can attain right relic abundance without resonance region. Nevertheless, this scenario can cause large scattering cross section between DM and nucleon in detector of direct searches, which is not flavored by current experimental limits.

\begin{figure}[H]
	\bc
	\includegraphics[scale=0.4]{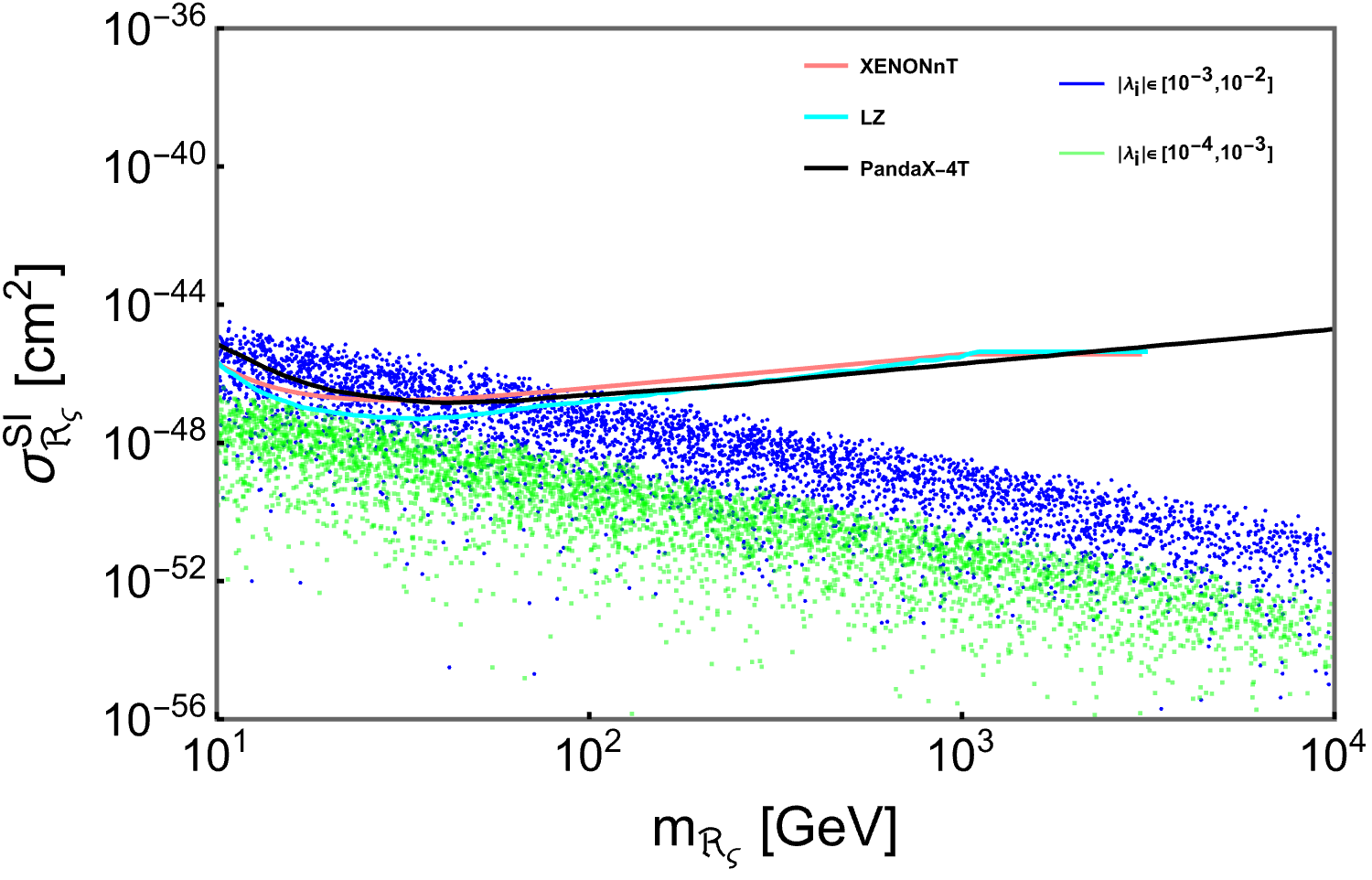}
	\caption[]{\label{scalarDM-DD} The dependence of predicted SI cross section $\sigma^{\text{SI}}_{\Re_{\varsigma}}$ as the function of its mass $m_{\Re_{\varsigma}}$. The pink, cyan and black solid lines present the current upper constraints of SI cross section reported by XENONnT \cite{XENON:2025vwd}, LZ \cite{LZ:2022lsv} and PandaX-4T\cite{PandaX:2024qfu} experiments. }
	\ec
\end{figure}

We want to stress that the effect of annihilation of scalar DM $\Re_{\varsigma}$ via new Higgs $H_1$ is disappeared in Fig. \ref{scalarDM} by suppressed vertex $\la_{\Re_{\varsigma}\Re_{\varsigma}H_1}\ll 1$ .  In our setup, $H_1$ is mostly composed by doublet $\eta$, thus scalar interaction term $\la_9(\eta^{\dagger}\eta)\varsigma\varsigma$ plays crucial role for $H_1$ portal annihilation of DM. For small $\la_9\sim \mathcal{O}(10^{-4}-10^{-2})$, the effect of $H_1$ portal is irrelevant, unless $\la_9$ is large enough, i.e. $\la_9\geq 0.1$. However, if $\la_9$ is large enough, the predicted cross section between DM and nucleon in direct searches can excess the current limit $\sigma^{\text{SI}}\simeq \mathcal{O}(10^{-48})$ cm$^2$. 
Direct detection constraints probe this scenario mainly through spin-independent scattering mediated by the Higgs boson, although additional contributions from $Z$,  $Z'$ exchange may also be present. Consequently, the scalar DM scenario is constrained by the interplay between relic density requirements, direct detection limits, and the structure of the scalar sector. The figure \ref{scalarDM-DD} illustrates the comparison between the predicted spin-independent (SI) cross section $\sigma^{\text{SI}}_{\Re_{\varsigma}}$ with current direct searches including XENONnT \cite{XENON:2025vwd}, LZ \cite{LZ:2022lsv} and PandaX-4T\cite{PandaX:2024qfu}.  We observe that for smaller interval $|\la_{i}|\in [10^{-4},10^{-3}]$, the model shows SI cross section is safe under the experimental limits, whereas for larger case $|\la_{i}|\in [10^{-3},10^{-2}]$, the lower bound of DM mass is required $m_{\Re_{\varsigma}}\geq 100$ GeV satisfying the these constraints. 

\subsubsection{Fermionic DM}

An alternative possibility is that the lightest \(Z_2\)-odd neutral state is fermionic, arising from the mixing of two singlet fermions of each kind. After diagonalization, the physical states \(\chi_{1,2}\) are obtained, with \(\chi_1\) identified as the dark matter candidate. The interactions of the mass eigenstates with the neutral gauge bosons can be written as
\begin{equation}
	\mathcal L_{\rm int}
	=
	g^{\prime\prime}\,
	\bar\chi_i \gamma^\mu (Q_D^{\rm mass})_{ij}\chi_j
	\left( \sin\alpha\, Z_\mu + \cos\alpha\, Z'_\mu \right),
\end{equation}
where \(Q_D^{\rm mass}\) encodes the mixing structure in the dark sector. The explicit form of the mass matrix, its diagonalization, and the resulting coupling matrix are presented in Appendix~\ref{app:fermionDM}. In this basis, the couplings of the DM candidate to the dark gauge boson \(Z'\), scalar mediators, and sterile neutrinos depend on the mixing angles \(\alpha\) and \(\theta_\chi\), which determine the relative strength of gauge and Yukawa interactions.

\begin{figure}[H]
	\bc
	\includegraphics[scale=0.4]{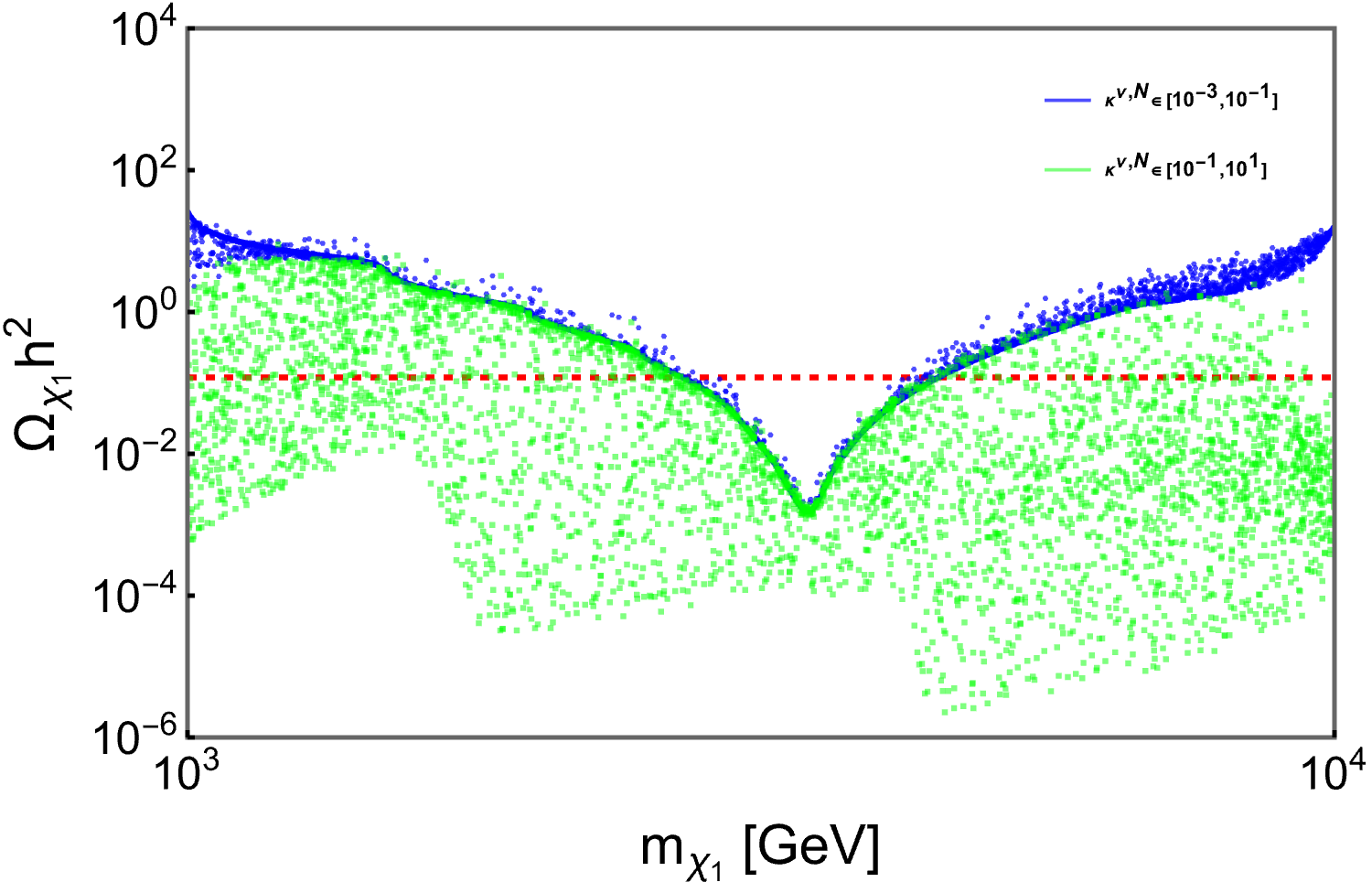}
	\caption[]{\label{fermionDM} The dependence of relic density $\Omega_{\chi_1}h^2$ as the function of its mass $m_{\chi_1}$. The dashed red line present the current experimental constraint of relic density $\Omega h^2\simeq 0.12$ \cite{Planck:2018vyg}. }
	\ec
\end{figure}

\begin{figure}[H]
	\bc
	\includegraphics[scale=0.4]{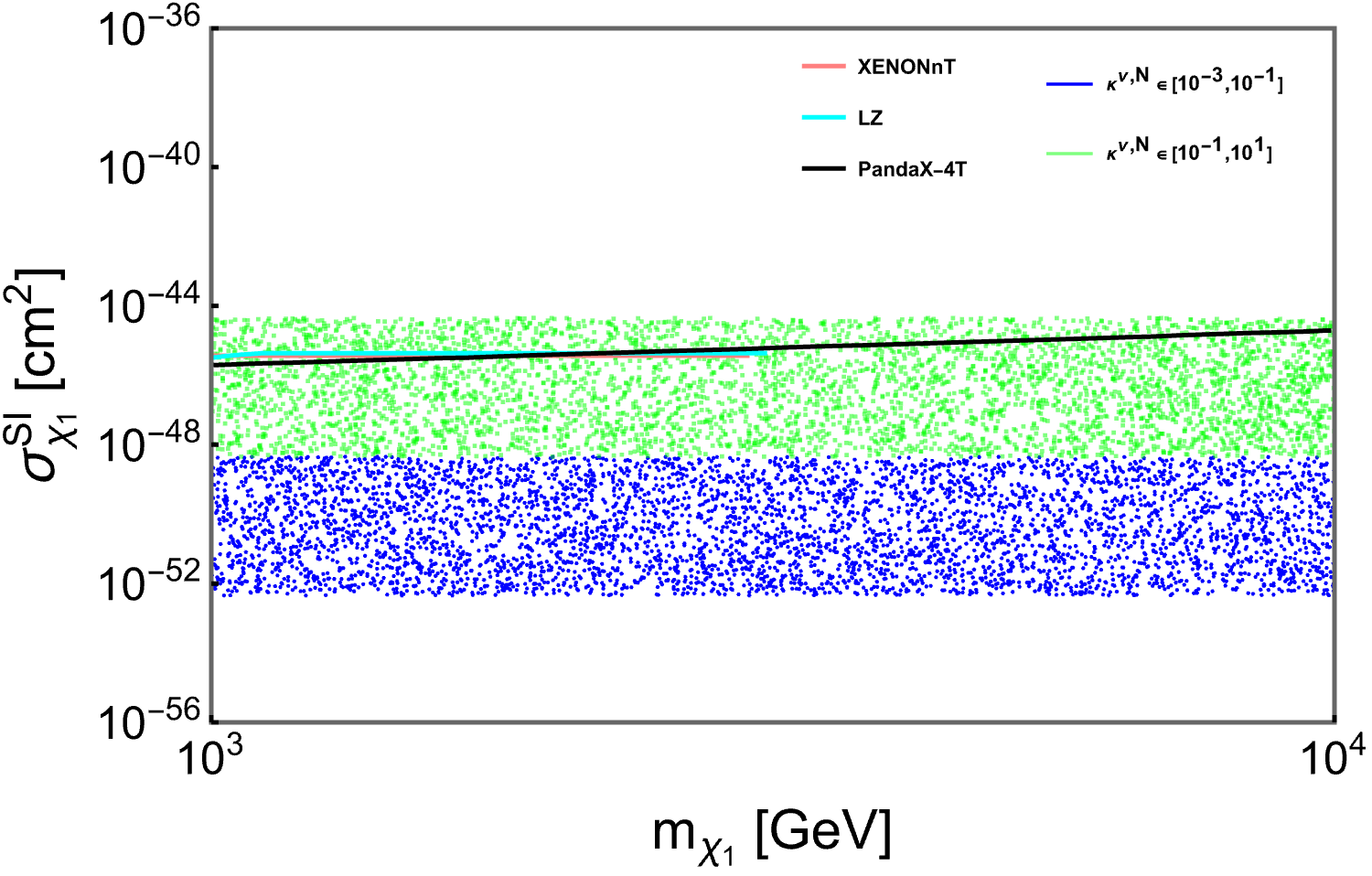}
	\caption[]{\label{fermionDM-DD} The dependence of predicted SI cross section $\sigma^{\text{SI}}_{\chi_1}$ as the function of its mass $m_{\chi_1}$. The pink, cyan and black solid lines present the current upper constraints of SI cross section reported by XENONnT \cite{XENON:2025vwd}, LZ \cite{LZ:2022lsv} and PandaX-4T\cite{PandaX:2024qfu} experiments.  }
	\ec
\end{figure}

In this case of fermionic DM, we fix the quartic couplings $\la_7=0.01,\la_7=0.02,\la_9=0.03,\la_{12}=-0.01$ since these parameters are irrelevant for the case of fermion DM. The figure \ref{fermionDM} displays the dependence of fermionic DM $\chi_1$ relic density as the function of its mass $m_{\chi_1}$ in two choices of Yukawa couplings $\ka^{\nu,N}\in[10^{-3},10^{-1}]$ (blue points) and $\ka^{\nu,N}\in[10^{-1},10^{1}]$ (green points). We see that there exists a resonance at $m_{\chi_1}\simeq m_{Z'}/2$,  in both cases of the Yukawa couplings. This is explained due to the fact that the dominant annihilation channels proceed via \(Z'\)-mediated processes.  Besides, the larger range of Yukawa couplings scenario $\ka^{\nu,N}\in[10^{-1},10^{1}]$ shows the more underabundant relic density compared to smaller Yukawa couplings $\ka^{\nu,N}\in [10^{-3},10^{-1}]$.  The relic abundance therefore depends on the DM mass, the dark gauge coupling, the mixing angles, and the masses of the mediator states. In particular, the resonant annihilation near \(m_{\chi_1} \simeq m_{Z'}/2\) or near scalar thresholds can play an important role in reproducing the observed DM relic density.
Direct detection signatures are model dependent and may arise from scalar-mediated interactions induced by Higgs mixing, or from gauge-mediated interactions involving the mixing of  \(Z, Z'\) bosons. In Fig.\ref{fermionDM-DD}, we demonstrate the dependence of SI cross section of fermionic DM $\chi_1$ as the function of its mass $m_{\chi_1}$ in two different cases of Yukawa couplings $\ka^{\nu,N}\in[10^{-3},10^{-1}]$ (blue points) and $\ka^{\nu,N}\in[10^{-1},10^{1}]$ (green points). For the larger Yukawa couplings, the SI cross section is shown to be larger than experimental constraints, while the smaller Yukawa couplings scenario fits the constraints better. 

\section{Conclusion}
\label{conclusion}
We have presented a minimal and predictive framework based on a dark $U(1)_D$ gauge symmetry that simultaneously addresses the origin of neutrino masses and DM within an inverse seesaw realization. The construction requires only a limited set of additional fields beyond the SM, providing an economical extension in which both sectors are naturally connected through a common gauge structure.

A central aspect of the model is the origin of the lepton-number-violating parameter $\mu$, which controls the smallness of light neutrino masses. Rather than being introduced by hand, $\mu$ emerges dynamically from the underlying theory. In the limit $\mu \to 0$, lepton number symmetry is restored, ensuring the technical naturalness of this parameter in the sense of ’t Hooft. This mechanism offers a theoretically well-motivated explanation for the suppressed scale of $\mu$, linking it directly to the dynamics of the extended sector.

We have derived the neutrino mass matrix and examined the resulting active--sterile mixing, which plays a crucial role in inducing deviations from unitarity as well as charged lepton flavor violation. The interplay between these effects has been analyzed in detail, showing that the model accommodates current experimental constraints while allowing for potentially observable signals in processes such as $\mu \to e \gamma$.

The dark $U(1)_D$ symmetry not only stabilizes the DM candidate but also establishes a direct connection between the neutrino and dark sectors. We have identified viable regions of parameter space consistent with relic density requirements, direct detection bounds, and collider constraints. In addition, the scalar sector leads to mixing with the SM Higgs boson, opening the possibility of observable deviations in Higgs phenomenology.

In summary, this framework provides a minimal and theoretically well-motivated realization of neutrino mass generation with a naturally small $\mu$ parameter, while simultaneously accounting for DM within a unified gauge structure. Future improvements in charged lepton flavor violation searches, non-unitarity probes, and DM experiments will be essential in testing this scenario and further constraining its parameter space.
\appendix
\section{Casas--Ibarra parametrization in the inverse seesaw}
\label{app:CI_inverse}

In this appendix, we briefly outline the derivation of the Casas--Ibarra parametrization adapted to the inverse seesaw framework.

In the basis $(\nu_L, \nu_R^c, N_R^c)$, the neutral fermion mass matrix takes the form
\begin{equation}
	\mathcal{M} =
	\begin{pmatrix}
		0 & m_D & 0 \\
		m_D^T & 0 & M \\
		0 & M^T & \mu
	\end{pmatrix}.
\end{equation}
In the limit $\mu \ll m_D \ll M$, the effective light neutrino mass matrix is given by
\begin{equation}
	m_\nu \simeq m_D M^{-1} \, \mu \, (M^T)^{-1} m_D^T.
	\label{eq:mnu_inverse}
\end{equation}

Defining the active--sterile mixing matrix as
\begin{equation}
	\Theta \simeq m_D M^{-1},
\end{equation}
Eq.~\eqref{eq:mnu_inverse} can be rewritten as
\begin{equation}
	m_\nu \simeq \Theta \, \mu \, \Theta^T.
	\label{eq:mnu_theta}
\end{equation}

The light neutrino mass matrix can be diagonalized as
\begin{equation}
	U_{\rm PMNS}^T \, m_\nu \, U_{\rm PMNS} = m_\nu^{\rm diag}.
\end{equation}
Substituting Eq.~\eqref{eq:mnu_theta}, one obtains
\begin{equation}
	U_{\rm PMNS}^T \, \Theta \, \mu \, \Theta^T \, U_{\rm PMNS}
	= m_\nu^{\rm diag}.
\end{equation}

Introducing the matrix
\begin{equation}
	X \equiv \mu^{1/2} \Theta^T U_{\rm PMNS},
\end{equation}
the above relation becomes
\begin{equation}
	X X^T = m_\nu^{\rm diag}.
\end{equation}
This equation admits the general solution
\begin{equation}
	X = R \, \sqrt{m_\nu^{\rm diag}},
\end{equation}
where $R$ is a complex orthogonal matrix satisfying
\begin{equation}
	R^T R = \mathbf{1}.
\end{equation}

Solving for $\Theta$, one finally obtains
\begin{equation}
	\Theta = U_{\rm PMNS} \, \sqrt{m_\nu^{\rm diag}} \, R \, \mu^{-1/2}.
	\label{eq:Theta_CI}
\end{equation}
This expression generalizes the Casas--Ibarra parametrization~\cite{Casas:2001sr} to the inverse seesaw scenario, where the small lepton-number-violating parameter $\mu$ replaces the role of the heavy mass scale in the conventional type-I seesaw. 

\section{Fermionic dark matter mixing and gauge interactions}
\label{app:fermionDM}

For simplicity, we consider two SM-singlet fermions \(\chi_R\) and \(\chi'_R\), carrying \(U(1)_D\) charges \(+1\) and \(-1\), respectively. In the interaction basis \((\chi_R,\chi'_R)\), the Majorana mass terms take the form,
\begin{equation}
	\mathcal{L}_{\rm mass} \supset 
	\frac{1}{2}
	\begin{pmatrix}
		\overline{\chi_R^c} & \overline{\chi_R^{\prime c}}
	\end{pmatrix}
	\begin{pmatrix}
		m_\chi & m_{\chi\chi'} \\
		m_{\chi\chi'} & m_{\chi'}
	\end{pmatrix}
	\begin{pmatrix}
		\chi_R \\
		\chi'_R
	\end{pmatrix}
	+ \text{h.c.}.
\end{equation}

The mass matrix is diagonalized by an orthogonal transformation
\begin{equation}
	U^T \mathcal{M}_\chi\, U = \mathrm{diag}(m_{\chi_1},\, m_{\chi_2}),
\end{equation}
with
\begin{equation}
	U=
	\begin{pmatrix}
		\cos\theta_\chi & \sin\theta_\chi \\
		-\sin\theta_\chi & \cos\theta_\chi
	\end{pmatrix},
	\qquad
	\tan 2\theta_\chi = \frac{2 m_{\chi\chi'}}{m_{\chi'} - m_\chi}.
\end{equation}
The mass eigenstates are defined as \((\chi_1,\chi_2)^T = U^T(\chi_R,\chi'_R)^T\).

\medskip

The interaction with the dark gauge boson \(C_\mu\) is given by
\begin{equation}
	\mathcal L_{\rm int}
	=
	g^{\prime\prime} C_\mu
	\begin{pmatrix}
		\bar\chi & \bar\chi'
	\end{pmatrix}
	\gamma^\mu
	\begin{pmatrix}
		1 & 0\\
		0 & -1
	\end{pmatrix}
	\begin{pmatrix}
		\chi\\
		\chi'
	\end{pmatrix}.
\end{equation}
Transforming to the mass basis yields
\begin{equation}
	\mathcal L_{\rm int}
	=
	g^{\prime\prime} C_\mu\,
	\bar\chi_i \gamma^\mu (Q_D^{\rm mass})_{ij}\chi_j,
\end{equation}
where
\begin{equation}
	Q_D^{\rm mass} = U^T
	\begin{pmatrix}
		1 & 0\\
		0 & -1
	\end{pmatrix}
	U
	=
	\begin{pmatrix}
		\cos 2\theta_\chi & \sin 2\theta_\chi\\
		\sin 2\theta_\chi & -\cos 2\theta_\chi
	\end{pmatrix}.
\end{equation}

After diagonalization of the neutral gauge boson sector, the dark gauge field is expressed as
\begin{equation}
	C_\mu = \sin\alpha\, Z_\mu + \cos\alpha\, Z'_\mu,
\end{equation}
leading to
\begin{align}
	\mathcal L_{\rm int}
	=\;&
	g^{\prime\prime}\sin\alpha\, Z_\mu\, \bar\chi_i \gamma^\mu (Q_D^{\rm mass})_{ij}\chi_j
	\nonumber\\
	&+
	g^{\prime\prime}\cos\alpha\, Z'_\mu\, \bar\chi_i \gamma^\mu (Q_D^{\rm mass})_{ij}\chi_j.
\end{align}

The diagonal and off-diagonal couplings are proportional to \(\cos 2\theta_\chi\) and \(\sin 2\theta_\chi\), respectively. In the limit of maximal mixing, \(\theta_\chi = \pi/4\), the diagonal interactions vanish and the couplings become purely off-diagonal.

\section*{\label{acknowledgement}Acknowledgement}
This research is funded by Vietnam National Foundation for Science and Technology Development (NAFOSTED) under grant number 103.01-2023.50.

\bibliographystyle{JHEP}
\bibliography{ISSU1D_Refs}
\end{document}